\documentclass[11pt]{article}

\pdfoutput=1

\usepackage{stackrel}
\usepackage{mathrsfs}
\usepackage{fancyref}
\usepackage{graphicx}
\usepackage{enumerate}
\usepackage{mdwlist}
\usepackage{caption}
\usepackage{fancybox}
\usepackage{pifont}
\usepackage[all]{xy}
\usepackage{makecell}
\usepackage{epigraph}
\usepackage{pifont}
\usepackage{dsfont}
\usepackage{relsize}
\usepackage{enumitem}
\usepackage{float}

\usepackage{epsfig}
\usepackage{graphicx}

\usepackage{latexsym,times}
\usepackage{amssymb,amsmath}
\usepackage{amsxtra}
\usepackage{amscd}
\usepackage{amsthm}
\usepackage{amsfonts}
\usepackage{dsfont}
\usepackage{xcolor}
\usepackage{cancel}
\usepackage{framed}

\usepackage{esint}

\usepackage{enumitem}
\usepackage{float}

\usepackage{hyperref}
\usepackage[square,sort&compress,comma,numbers]{natbib}

\makeatletter

\@addtoreset{equation}{section} \makeatother

\newcommand{\half}{{{\textstyle\frac{1}{2}}}}
\newcommand{\quarter}{{{\textstyle\frac{1}{4}}}}
\newcommand{\be}{\begin{equation}}
\newcommand{\ee}{\end{equation} }
\newcommand{\beqa}{\begin{eqnarray} }
\newcommand{\eeqa}{\end{eqnarray} }
\newcommand{\ba}{\begin{array}}
\newcommand{\ea}{\end{array}}
\newcommand{\bpm}{\begin{pmatrix}}
\newcommand{\epm}{\end{pmatrix}}

\newcommand{\Spin}{\mathbf{Spin}}

\newcommand{\ODD}{\mathbf{O}(D,D)}

\newcommand{\hODD}{\mathbf{O}(\hD,\hD)}

\newcommand{\ODDpr}{\mathbf{O}(\Dpr,\Dpr)}

\def\Apr{A^{\prime}}
\def\Bpr{B^{\prime}}

\def\Dpr{D^{\prime}}

\def\Lpr{L^{\prime}}
\def\Mpr{M^{\prime}}
\def\Npr{N^{\prime}}

\def\cJpr{\cJ^{\prime}}

\newcommand\rd{{\rm d}}
\newcommand\rD{{\rm D}}

\newcommand\cA{{\cal A}}

\newcommand\cC{{\cal C}}
\newcommand\cD{{\cal D}}
\newcommand\cE{{\cal E}}

\newcommand\cH{{\cal H}}

\newcommand\cJ{{\cal J}}

\newcommand\cL{{\cal L}}

\newcommand\cX{{\cal X}}
\newcommand\cY{{\cal Y}}

\newcommand\hcJ{{\hat{\cal J}}}
\newcommand\hcL{{\hat{\cal L}}}

\newcommand\hmu{\hat{\mu}}
\newcommand\hnu{\hat{\nu}}

\newcommand\hA{\hat{A}}
\newcommand\hB{\hat{B}}

\newcommand\hcH{\hat{\cH}}
\newcommand\hD{\hat{D}}

\newcommand\hM{{\hat{M}}}
\newcommand\hN{{\hat{N}}}

\newcommand\hW{{\hat{W}}}

\newcommand\cHpr{\cH^{\prime}}

\newcommand\dis{\displaystyle}

\def\tx{\tilde{x}}
\def\ty{\tilde{y}}
\def\tpartial{\tilde{\partial}}

\def\hati{\hat{\imath}}

\def\bra{\bar{a}}
\def\brb{\bar{b}}

\def\breta{\bar{\eta}}

\def\brh{\bar{h}}
\def\bri{\bar{\imath}}
\def\brj{\bar{\jmath}}
\def\brk{{\bar{k}}}
\def\brl{{\bar{l}}}
\def\brm{{\bar{m}}}
\def\brn{{\bar{n}}}
\def\brp{{\bar{p}}}
\def\brq{{\bar{q}}}

\def\brB{\bar{B}}

\def\brR{\bar{R}}

\def\brV{{\bar{V}}}
\def\brW{{\bar{W}}}
\def\brX{{\bar{X}}}
\def\brY{{\bar{Y}}}
\def\brP{{\bar{P}}}

\def\brcX{{\bar{\cX}}}
\def\brcY{{\bar{\cY}}}

\newcommand{\na}{{\nabla}}

\newcommand{\bea}{\begin{eqnarray}}
\newcommand{\eea}{\end{eqnarray}}

\newcommand{\Ker}{\text{\rm Ker }}
\newcommand{\p}{\partial}

\newcommand{\pl}{\left(}
\newcommand{\pr}{\right)}
\definecolor{rougef}{rgb}{0.56,0,0}		
\definecolor{vertf}{rgb}{0,0.5,0}		
\definecolor{bleuf}{rgb}{0,0,0.8}

\newcommand{\ie}{\textit{i.e.}~}
\newcommand{\cf}{\textit{cf.}~}
\newcommand{\eg}{\textit{e.g.}~}

\begin{document}

\begin{titlepage}

\title{\vskip -100pt
\vskip 2cm 
{{{Classification of non-Riemannian  doubled-yet-gauged spacetime
\\~}}}}

\author{\sc Kevin Morand${}^{\sharp\, \flat}$  \quad and\quad Jeong-Hyuck Park${}^{\dagger\ast}$}
\date{}
\maketitle 
\begin{center}
${}^{\sharp}$Departamento de Ciencias F\'isicas, Universidad Andres Bello
\\ Republica 220, Santiago de Chile, Chile\\
\vspace{2mm}
${}^{\flat}$Departamento de F\'isica, Universidad T\'ecnica Federico Santa Mar\'ia\\
Centro Cient\'ifico-Tecnol\'ogico de Valpara\'iso, Casilla 110-V, Valpara\'iso, Chile\\
~\\

${}^{\dagger}$Department of Physics, Sogang University, 35 Baekbeom-ro, Mapo-gu,  Seoul  04107, Korea\\
\vspace{2mm}
${}^{\ast}$Center for Theoretical Physics of the Universe, Institute for Basic Science (IBS), Seoul 08826,  Korea\\
~\\
\texttt{kevin.morand@lmpt.univ-tours.fr\qquad park@sogang.ac.kr }\\
~\\
~\\

\end{center}
\begin{abstract}
\vskip0.1cm  
\noindent   Assuming  $\mathbf{O}(D,D)$ covariant fields as the   `fundamental' variables, Double Field Theory can  accommodate  novel geometries  where a Riemannian metric cannot be defined, even   locally.   Here we present a complete classification of    such non-Riemannian  spacetimes  in terms of  two non-negative integers, $(n,\bar{n})$, $0\leq n+\bar{n}\leq D$.  Upon these backgrounds,  strings become chiral and anti-chiral  over $n$ and $\bar{n}$ directions respectively, while  particles and  strings are frozen over the $n+\bar{n}$ directions.  In particular,  we identify  $(0,0)$ as  Riemannian manifolds, $(1,0)$ as non-relativistic spacetime,  $(1,1)$ as   Gomis-Ooguri   non-relativistic string, $(D{-1},0)$ as ultra-relativistic  Carroll geometry, 
 and $(D,0)$ as  Siegel's chiral string. Combined with a covariant Kaluza-Klein ansatz which we further spell, $(0,1)$  leads to Newton-Cartan   gravity. 
  Alternative to the  conventional  string  compactifications    on small manifolds,  non-Riemannian spacetime such as $D=10$, $(3,3)$  may open  a  new  scheme of  the  dimensional reduction from ten to four.

\end{abstract}

\thispagestyle{empty}

\end{titlepage}

\tableofcontents

\section{Introduction}
Ever since  Einstein formulated his theory of gravity, \textit{i.e.~}General Relativity (GR), by employing differential geometry \textit{\`a la} Riemann,      the  Riemannian    metric, $g_{\mu\nu}$, has been privileged to be   the only  {geometric}   and thus  {gravitational} field.  All other fields    are meant to  be     `extra matters'.   On the other hand,   string theory  suggests  us to put a  two-form gauge potential, $B_{\mu\nu}$, and a scalar dilaton, $\phi$, on an equal footing along with the metric. Forming the  massless sector of closed strings, this triplet of objects is ubiquitous in all  string theories. 
Further, a genuine stringy  symmetry, T-duality,  can  mix the three of them~\cite{Buscher:1987sk,Buscher:1987qj}, thus hinting at the existence of    \textit{Stringy Gravity} which should take  the entire   closed string massless sector  as geometric  and  gravitational.  After series of pioneering works on `doubled sigma models'~\cite{Duff:1989tf,Tseytlin:1990nb,Tseytlin:1990va,Hull:2004in,Hull:2006qs,Hull:2006va} and `double field theory' (DFT)~\cite{Siegel:1993xq,Siegel:1993th,Hull:2009mi,Hull:2009zb,Hohm:2010pp} (\cf \cite{Aldazabal:2013sca,Berman:2013eva,Hohm:2013bwa} for reviews),  such an idea of Stringy Gravity has    materialized.\footnote{Strictly speaking, string theory does not predict General Relativity but its own gravity, \ie Stringy Gravity.   }

\noindent The word  `double' above refers to the fact that  doubled $(D\,{+D})$-dimensional coordinates are  used for the description of  $D$-dimensional physical  spacetime. While such a usage was  historically first made   in the case of a torus background --    by introducing a dual coordinate conjugate to the  string   winding momentum --  the  doubled coordinates are    far more general and can be applied to any compact or  non-compact spacetime, and to   not only string but also   particle theories.  

~\\
\noindent Stringy Gravity of our interest  adopts    the \textit{doubled-yet-gauged coordinate system}~\cite{Park:2013mpa} which meets   two  properties.  Firstly,   an  $\ODD$ group is  \textit{a priori}   postulated,   having the  invariant constant ``metric",   
\be
\cJ_{AB}=\left(\ba{cc}{0}&{1}\\{1}&{0}\ea\right)\,.
\label{cJ}
\ee
 Along with its inverse,  $\cJ^{AB}$, the invariant metric can be used to freely raise and lower the $\ODD$ vector indices (capital letters, $A,B,\ldots$). Secondly,     the doubled coordinates are   gauged by  an   equivalence relation,
\be
x^{A}~\sim~x^{A}+\Delta^{A}(x)\,,
\ee
where $\Delta^{A}$ is an arbitrary `derivative-index-valued'  vector. This   means that its superscript index must be identifiable  as that of a  derivative, $\partial^{A}=\cJ^{AB}\partial_{B}$. For example, with arbitrary functions, ${\Phi}_{1}$, $\Phi_{2}$ belonging to the theory,  $\Delta^{A}=\Phi_{1}\partial^{A}\Phi_{2}$.  
The equivalence relation can be    realized by  requiring that all the fields or functions in Stringy Gravity -- such as $\Phi_{1}$, $\Phi_{2}$,  physical fields,  local symmetry parameters,   and  their arbitrary  derivatives --  should be  invariant under the coordinate gauge symmetry shift,
\be
\Phi(x+\Delta)=\Phi(x)\quad\Longleftrightarrow\quad\Delta^{A}\partial_{A}=0\,.
\label{CGS}
\ee 
In this way, a single physical point is not represented by a point, as in ordinary Riemannian geometry,  but  as a gauge orbit  in the doubled coordinate system.

\noindent The above coordinate gauge symmetry  invariance is equivalent to the so-called `section condition' in DFT, \be
\partial_{A}\partial^{A}=0\,.
\label{SECCON}
\ee 
With respect to the off block-diagonal form of the $\ODD$  metric~(\ref{cJ}), the doubled coordinates split into two parts: $x^{A}=(\tilde{x}_{\mu}, x^{\nu})$ and $\partial_{A}=(\tpartial^{\mu},\partial_{\nu})$, such that $\partial_{A}\partial^{A}=2\partial_{\mu}\tpartial^{\mu}$.  The general solution to the section condition is then   given by $\tpartial^{\mu}\equiv 0$, up to $\ODD$ rotations~\cite{Siegel:1993th,Hull:2009mi}. 

\noindent Diffeomorphism covariance in doubled-yet-gauged spacetime reads
\be
\ba{cc}
\delta x^{A}=\xi^{A}\,,
\quad&\quad
\delta\partial_{A}=-\partial_{A}\xi^{B}\partial_{B}=(\partial^{B}\xi_{A}-\partial_{A}\xi^{B})\partial_{B}\,,
\ea
\label{diff1}
\ee
and  for a covariant tensor (or tensor density with weight $\omega$),
\be
\delta T_{A_{1}\cdots A_{n}}=-\omega\partial_{B}\xi^{B}T_{A_{1}\cdots A_{n}}+\textstyle{\sum_{i=1}^{n}\,}(\partial_{B}\xi_{A_{i}}-\partial_{A_{i}}\xi_{B})T_{A_{1}\cdots A_{i-1}}{}^{B}{}_{A_{i+1}\cdots  A_{n}}\,.
\label{diff2}
\ee
The latter corresponds to the passive counterpart  of  the ``generalized Lie derivative", $\hcL_{\xi}$, \textit{\`a la} Siegel~\cite{Siegel:1993th}.

\noindent The whole  massless sector of closed strings, or \textit{stringy gravitons},  can be     represented by  a unit-weighted  scalar density, $e^{-2d}$, and a symmetric projector,   
\be
\ba{ll}
 P_{AB}=P_{BA}\,,\qquad&\qquad P_{A}{}^{B}P_{B}{}^{C}=P_{A}{}^{C}\,.
\ea
\label{Pdef}
\ee 
The  complementary,  orthogonal projector, $\brP_{AB}=\cJ_{AB}-P_{AB}$, satisfies, from (\ref{Pdef}), $P\brP=0$, $\brP^{2}=\brP$.  Covariant derivatives, $\na_{A}=\partial_{A}+\Gamma_{A}$,  scalar curvature $S_{\scriptscriptstyle{(0)}}$ and ``Ricci-like''  curvature $P_{A}{}^{C}\brP_{B}{}^{D}S_{CD}$  are then expressed in terms of $\{P_{AB}, \brP_{AB}, d\}$ and their derivatives or equivalently in terms of the stringy analog  of the Christoffel symbol, $\Gamma_{ABC}$~\eqref{Gammao}~\cite{Jeon:2011cn}.\footnote{However, a fully covariant four-indexed ``Riemann-like''   curvature has been argued not to exist ~\cite{Jeon:2011cn,Hohm:2011si}. This absence is, in a way, consistent with the fact that there exists no locally inertial frame for an extended object, \ie a string,  where the stringy Christoffel connection~\eqref{Gammao} might vanish completely: Equivalence Principle holds for particles not strings.}

  The difference of the two projectors  sets a symmetric $\ODD$ element, known as the DFT-metric (or ``generalized metric"),  
\be
\cH_{AB}=\cH_{BA}=P_{AB}-\brP_{AB}\qquad\text{ satisfying }\qquad
\cH_{A}{}^{C}\cH_{B}{}^{D}\cJ_{CD}=\cJ_{AB}\,.
\label{cHdef0}
\ee 
These $\ODD$ covariant  defining properties of the  stringy gravitational  fields can be   conveniently solved   by the conventional variables,  
\be
\ba{ll}
\cH_{MN}=\left(\ba{cc}g^{\mu\nu}&-g^{\mu\sigma}B_{\sigma\lambda}\\
B_{\kappa\rho}g^{\rho\nu}&~~g_{\kappa\lambda}-B_{\kappa\rho}g^{\rho\sigma}B_{\sigma\lambda}\ea\right)\,.
\ea
\label{RiemanncH}
\ee
However, this is not the most general solution: counter examples have been reported where the upper left $D\times D$ block of $\cH_{AB}$ is degenerate~\cite{Lee:2013hma,Ko:2015rha,Park:2016sbw},  and  have been  shown to provide  a natural geometry for the non-relativistic string theory {\it{\`a la}} Gomis and Ooguri~\cite{Gomis:2000bd}.\footnote{\cf\cite{Malek:2013sp,Blair:2013gqa} for 
 U-duality analogs.}  Namely, by assuming the $\ODD$ covariant variables as \textit{the} fundamental  fields, DFT or  Stringy Gravity  becomes more general than GR: it encompasses   `non-Riemannian'  spacetime    where  the Riemannian  metric, $g_{\mu\nu}$,   cannot be   defined,  even locally. This includes various `singular' limits of the Riemannian metric of which  the inverse, $g^{\mu\nu}$, becomes degenerate (\textit{c.f.~}T-fold,  `non-geometries' or `waves' in  the global sense~\cite{Hull:2004in,Hull:2006qs,Hull:2006va,Blumenhagen:2011ph,
Blumenhagen:2012nt,Dibitetto:2012rk,Cederwall:2014opa,Berkeley:2014nza,Berman:2014jsa,Lee:2016qwn}). \\

\paragraph{Scope of the paper}
~\\
It is the purpose of the present paper  to classify  completely   DFT backgrounds, by deriving the most general solution to the defining property of the stringy gravitational field,  or (\ref{cHdef0}). Our classification is given in terms of two non-negative integers,  $(n,\brn)$, $0\leq n+\bar{n}\leq D$.  Except for the $(0,0)$ case,  these are generically non-Riemannian.

\noindent Since various DFTs  and  the relevant  doubled sigma models  have been  constructed,  strictly    in terms of the  $\ODD$ covariant fields without referring to the Riemannian ones $\{g_{\mu\nu},B_{\mu\nu},\phi\}$,\footnote{Yet, there are quite a few works in the literature which do not meet this criterion, relying explicitly on the Riemannian variables. Our results  are thus not  applicable therein.}  our result can be \textit{readily and unambiguously} applicable to these models which include, \textit{e.g.}~coupling to the Standard Model~\cite{Choi:2015bga},   higher spin~\cite{Bekaert:2016isw}, fluctuation or Noether  analyses~\cite{Ko:2015rha,Hohm:2015ugy,Park:2015bza,Blair:2015eba}, the doubled-yet-gauged Green-Schwarz superstring action~\cite{Park:2016sbw} and the maximally supersymmetric $D=10$ DFT~\cite{Jeon:2012hp}. In particular, this last example,  with the Killing spinor equations therein,  may lead to   a  new  scheme of  the  dimensional reduction from ten to four, by assuming  the  six-dimensional internal space to be non-Riemannian,  
alternative to the traditional   string  compactification   on `small' Riemannian  manifolds~\cite{Grana:2005jc}. 
 Further applications can be found in the  holographic correspondences  between   Newton-Cartan gravity and  condensed matter physics, \textit{e.g.~}\cite{Son:2013rqa,BergshoeffSimons}, as well as in the context of non-relativistic strings \cite{Andringa:2012uz,Ko:2015rha,Park:2016sbw,Harmark:2017rpg}. 
~\\

\paragraph{Organization of the paper}
~\\
The rest of the paper is structured as follows. In section~\ref{SECG}, we classify  the DFT-metric and spell the corresponding  DFT-vielbeins. We  discuss the  dynamics of  point particle and string upon these backgrounds. We also spell a path integral definition of the proper length in doubled-yet-gauged spacetime as well as  a covariant Kaluza-Klein  ansatz for DFT.     In section~\ref{SECA}, we discuss various applications, such as Gomis-Ooguri  non-relativistic string,  non-relativistic and ultra-relativistic geometries, Siegel's chiral string and Newton-Cartan gravity. Appendix contains the technical derivation of the main result.  

\newpage

\section{General results\label{SECG}}

\subsection{Classification of the DFT-metric}

As recalled in the introduction, the DFT-metric is  by definition a symmetric $\ODD$ element,  satisfying  the following relation:
\be
\ba{cc}
\cH_{AB}=\cH_{BA}\,,\quad&\quad\cH_{A}{}^{B}\cH_{B}{}^{C}=\delta_{A}{}^{C}\,.
\ea
\label{cHdef}
\ee 
Our main result consists in providing a full classification for DFT-metrics by solving the above defining properties: the most general solution is characterized   by two  non-negative integers, $(n,{\brn})$,  $0\leq n+\brn\leq D$, and assumes the following  form:
\be
\cH_{AB}=\left(\ba{cc}H^{\mu\nu}&
-H^{\mu\sigma}B_{\sigma\lambda}+Y_{i}^{\mu}X^{i}_{\lambda}-
\brY_{\bri}^{\mu}\brX^{\bri}_{\lambda}\\
B_{\kappa\rho}H^{\rho\nu}+X^{i}_{\kappa}Y_{i}^{\nu}
-\brX^{\bri}_{\kappa}\brY_{\bri}^{\nu}\quad&~~
~~K_{\kappa\lambda}-B_{\kappa\rho}H^{\rho\sigma}B_{\sigma\lambda}
+2X^{i}_{(\kappa}B_{\lambda)\rho}Y_{i}^{\rho}
-2\brX^{\bri}_{(\kappa}B_{\lambda)\rho}\brY_{\bri}^{\rho}
\ea\right)\,
\label{cHFINAL}
\ee
where $i,j=1,2,\cdots, n$ and  $\bri,\brj=1,2,\cdots,  \brn$. The variables, $\left\{H^{\mu\nu},\,K_{\mu\nu}\,, \,B_{\mu\nu}\,,\,X^{i}_{\mu}\,,\,Y^{\nu}_{j}\,,\,\brX^{\bri}_{\mu}\,,\,\brY_{\brj}^{\nu}\right\}$,  must meet the following properties: 
\begin{itemize}
\item[--]  $H^{\mu\nu}$ and $K_{\mu\nu}$ are symmetric tensors 
\be
\ba{llll}
H^{\mu\nu}=H^{\nu\mu}\,,\quad&\quad K_{\mu\nu}=K_{\nu\mu}\, , 
\ea
\label{HK}
\ee
whose kernels are spanned by $\big\{X^{i}_{\mu},\brX^{\bri}_{\nu}\big\}$  and $\big\{Y_{j}^{\mu},\brY^{\nu}_{\brj}\big\}$, respectively
\be
\ba{llll}
H^{\mu\nu}X^{i}_{\nu}=0\,,\quad&\quad
H^{\mu\nu}\brX^{\bri}_{\nu}=0\,;~\quad&~\quad
K_{\mu\nu}Y_{j}^{\nu}=0\,,\quad&\quad
K_{\mu\nu}\brY_{\brj}^{\nu}=0\,;
\ea
\label{HXX}
\ee
\item[--] a completeness relation,
\be
H^{\mu\rho}K_{\rho\nu}
+Y_{i}^{\mu}X^{i}_{\nu}+\brY_{\bri}^{\mu}\brX^{\bri}_{\nu}
=\delta^{\mu}{}_{\nu}\,;
\label{COMP}
\ee
\item[--] the skew-symmetry of  $B$-field,
\be
B_{\mu\nu}=-B_{\nu\mu}\,.
\label{Bskew}
\ee 
\end{itemize}

While the  derivation is carried out in  Appendix, 
some comments are in order.  From (\ref{HXX}), (\ref{COMP}) and the linear independency of $\left\{X^{i}_{\mu}, \brX^{\bri}_{\nu}\right\}$,  orthonormal as well as algebraic  relations  follow
\be
\ba{lllll}
Y^{\mu}_{i}X_{\mu}^{j}=\delta_{i}{}^{j}\,,\quad&\,
\brY^{\mu}_{\bri}\brX_{\mu}^{\brj}=\delta_{\bri}{}^{\brj}\,,\quad&\,
Y^{\mu}_{i}\brX_{\mu}^{\brj}=
\brY^{\mu}_{\bri}X_{\mu}^{j}=0\,,\quad&\,
H^{\rho\mu}K_{\mu\nu}H^{\nu\sigma}=H^{\rho\sigma}\,,\quad&\,
K_{\rho\mu}H^{\mu\nu}K_{\nu\sigma}=K_{\rho\sigma}\,.
\ea
\label{complete}
\ee
With the choice of the section,  $\tpartial^{\mu}\equiv 0$,  the doubled-yet-gauged  diffeomorphisms~(\ref{diff1}), (\ref{diff2}), or the generalized Lie derivative of the DFT-metric, \cf {\eqref{gLie}}, are compatible with the following transformations: 
\be
\ba{llll}
\quad\delta X^{i}_{\mu}=\cL_{\xi}X^{i}_{\mu}\,,\qquad&~~\quad
\delta \brX^{\bri}_{\mu}=\cL_{\xi}\brX^{\bri}_{\mu}\,,\qquad&~~\quad
\delta Y_{j}^{\nu}=\cL_{\xi}Y_{j}^{\nu}\,,\qquad&~~\quad
\delta \brY_{\brj}^{\nu}=\cL_{\xi}\brY_{\brj}^{\nu}\,,\\
\multicolumn{4}{c}{\delta H^{\mu\nu}=\cL_{\xi}H^{\mu\nu}\,,\qquad\quad
\delta K_{\mu\nu}=\cL_{\xi}K_{\mu\nu}\,,\qquad\quad
\delta B_{\mu\nu}=\cL_{\xi}B_{\mu\nu}+\partial_{\mu}\tilde{\xi}_{\nu}-\partial_{\nu}\tilde{\xi}_{\mu}\,,}
\ea
\label{deltaXYHK}
\ee
where $\cL_{\xi}$ denotes the ordinary, \textit{i.e.~}undoubled,  Lie derivative with the local parameter, $\xi^{\nu}$,  being part of the doubled vector field, $\,\xi^{A}=(\tilde{\xi}_{\mu},\xi^{\nu})$.  Our $(n,\brn)$-classification of the DFT-metric having the explicit parametrization~\eqref{cHFINAL} is particularly useful for the choice of the section, $\tpartial^{\mu}\equiv0$. For example, the action for  a massless scalar field reads  (\textit{c.f.~}\cite{Bergshoeff:2017btm})
\be
\int_{\scriptsize{\rm{section}}} e^{-2d}~\cH^{AB}\partial_{A}\Phi\partial_{B}\Phi~\equiv~
\int \rd^{D}x~ e^{-2d}~H^{\mu\nu}\partial_{\mu}\Phi\partial_{\nu}\Phi\,.
\label{Scalar}
\ee
For couplings to generic tensors or Yang-Mills fields, we refer  to \cite{Jeon:2011cn,Jeon:2010rw,Jeon:2011kp,Park:2015bza,Ko:2015rha}.   However, if the  $(n,\brn)$ DFT-metric~(\ref{cHFINAL})  admits an  isometry direction,  there appears  arbitrariness in  choosing  the section. In this case, our parametrization~(\ref{cHFINAL}) may be modified, see \textit{e.g.}~\cite{Andriot:2013xca,Lee:2016qwn}.

\noindent Clearly, constant $(n,\brn)$ DFT-metric~(\ref{cHFINAL}) and DFT-dilaton, $d$,  solve  the equations of motion of DFT. Thus, our $(n,\brn)$ classification also accounts for non-Riemannian `flat' backgrounds. 
It is worthwhile to note that the characteristic  value,  $(n,\brn)$, may change point-wise in a given doubled-yet-gauged curved spacetime, typically at a ``Riemannian singular point''. Further, $\ODD$ rotations (along  isometry directions) can also change the value of $(n,\brn)$,  for example, the $(0,0)$ fundamental string background \textit{\`a la} Dabholkar \textit{et al.}~\cite{Dabholkar:1990yf}  can be mapped to  $(1,1)$  by certain  $\ODD$ rotations~\cite{Lee:2013hma} (\cf\cite{Malek:2013sp}). However, the trace of a DFT-metric, 
\be
\cH_{A}{}^{A}=2(n-\brn)\,,
\label{trace}
\ee
remains invariant under $\ODD$ rotations  and further  point-wise  if we fix the underlying spin group~(\ref{cHtrace}).

It is instructive to note that the $B$-field contributes to the DFT-metric by an $\ODD$ conjugation, 
\be
\cH_{AB}=\left(\ba{cc}1&0\\B&1\ea\right)\left(\ba{cc}H~&~
Y_{i}(X^{i})^{T}-\brY_{\bri}(\brX^{\bri})^{T}\\
X^{i}(Y_{i})^{T}-\brX^{\bri}(\brY_{\bri})^{T}\quad~&~\quad K
\ea\right)\left(\ba{cc}1&-B\\0&1\ea\right)\,,
\label{Bcontri}
\ee
such that the contribution is  `Abelian', in the following sense:
\be
\left(\ba{cc}1&0\\B_{1}&1\ea\right)
\left(\ba{cc}1&0\\B_{2}&1\ea\right)
=\left(\ba{cc}1&0\\B_{1}+B_{2}&1\ea\right)\,.
\label{BAbelian}
\ee
Further,  the precise  expression  of the $(n,\brn)$ DFT-metric~(\ref{cHFINAL}) as well as the  fundamental algebraic relations, (\ref{HXX}), (\ref{COMP}), (\ref{Bskew}) are invariant under several transformations.  

Firstly under   obvious $\mathbf{GL}(n)\times\mathbf{GL}(\brn)$ rotations,
\be
\ba{ccc}
\left(X^{i}_{\mu}\,,\,Y_{i}^{\mu}\,,\,\brX^{\bri}_{\mu}\,,\,\brY_{\bri}^{\nu}\right)\quad&\longmapsto&\quad
\left(X^{j}_{\mu}\,R_{j}{}^{i}\,,\,R^{-1}{}_{i}{}^{j}\,Y_{j}^{\nu}\,,\,\brX^{\brj}_{\mu}\,\brR_{\brj}{}^{\bri}\,,\,\brR^{-1}{}_{\bri}{}^{\brj}\,\brY_{\brj}^{\nu}\right)\,;
\ea
\ee
secondly under the transformation of  only the $B$-field having two arbitrary  skew-symmetric local parameters, $m_{ij}=-m_{ji}$, $\brm_{\bri\brj}=-\brm_{\brj\bri}$,
\be
\ba{cll}
B_{\mu\nu}~~&\longmapsto&~~ B_{\mu\nu}+X_{\mu}^{i}X_{\nu}^{j}m_{ij}
+\brX_{\mu}^{\bri}\brX_{\nu}^{\brj}\brm_{\bri\brj}\,;
\ea
\label{Bonly}
\ee
and lastly  under the following somewhat less trivial transformations of $\{Y_{i}^{\mu}\,,\,\brY_{\bri}^{\mu}\,,\,K_{\mu\nu}\,,\,B_{\mu\nu}\}$,
\be
\ba{cll}
Y_{i}^{\mu}~~&\longmapsto&~~ Y_{i}^{\mu}+H^{\mu\nu}V_{\nu i}\,,\\
\brY_{\bri}^{\mu}~~&\longmapsto&~~\brY_{\bri}^{\mu}+H^{\mu\nu}\brV_{\nu\bri}\,,\\
K_{\mu\nu}~~&\longmapsto&~~ K_{\mu\nu}-2X^{i}_{(\mu}K_{\nu)\rho}H^{\rho\sigma}V_{\sigma i}-2\brX^{\bri}_{(\mu}K_{\nu)\rho}H^{\rho\sigma}\brV_{\sigma\bri}+(X_{\mu}^{i}V_{\rho i}+\brX_{\mu}^{\bri}\brV_{\rho\bri})H^{\rho\sigma}(X_{\nu}^{j}V_{\sigma j}+\brX_{\nu}^{\brj}\brV_{\sigma\brj})\,,\\
B_{\mu\nu}~~&\longmapsto&~~ B_{\mu\nu}-2X^{i}_{[\mu}K_{\nu]\rho}H^{\rho\sigma}V_{\sigma i}+2\brX^{\bri}_{[\mu}K_{\nu]\rho}H^{\rho\sigma}\brV_{\sigma\bri}
+2X^{i}_{[\mu}\brX^{\bri}_{\nu]}V_{\rho i}H^{\rho\sigma}\brV_{\sigma\bri}\,,
\ea
\label{MilneG}
\ee
where $V_{\mu i}$ and $\brV_{\mu\bri}$ are arbitrary local parameters.  In fact, the latter  two transformations, (\ref{Bonly}) and (\ref{MilneG}), can be unified into 
\be
\ba{cll}
Y_{i}^{\mu}~~&\longmapsto&~~ Y_{i}^{\mu}+H^{\mu\nu}V_{\nu i}\,,\\
\brY_{\bri}^{\mu}~~&\longmapsto&~~\brY_{\bri}^{\mu}+H^{\mu\nu}\brV_{\nu\bri}\,,\\
K_{\mu\nu}~~&\longmapsto&~~ K_{\mu\nu}-2X^{i}_{(\mu}K_{\nu)\rho}H^{\rho\sigma}V_{\sigma i}-2\brX^{\bri}_{(\mu}K_{\nu)\rho}H^{\rho\sigma}\brV_{\sigma\bri}+(X_{\mu}^{i}V_{\rho i}+\brX_{\mu}^{\bri}\brV_{\rho\bri})H^{\rho\sigma}(X_{\nu}^{j}V_{\sigma j}+\brX_{\nu}^{\brj}\brV_{\sigma\brj})\,,\\
B_{\mu\nu}~~&\longmapsto&~~ B_{\mu\nu}
-2X^{i}_{[\mu}V_{\nu]i}+2\brX^{\bri}_{[\mu}\brV_{\nu]\bri}
+2X^{i}_{[\mu}\brX^{\bri}_{\nu]}\left(Y_{i}^{\rho}\brV_{\rho\bri}
+\brY_{\bri}^{\rho}V_{\rho i}+V_{\rho i}H^{\rho\sigma}\brV_{\sigma\bri}\right)\,.
\ea
\label{MilneG2}
\ee
Note that  in (\ref{MilneG}) the local parameters appear only through the contractions with $H^{\mu\nu}$, \textit{i.e~}$H^{\mu\nu}V_{\nu i}$ and $H^{\mu\nu}\brV_{\nu i}$. On the other hand in (\ref{MilneG2}), the $B$-field transformation contains orthogonal contributions. Substituting $V_{\mu i}=-\half m_{ij}X^{j}_{\mu}$ and  $\brV_{\mu\bri}=\half\brm_{\bri\brj}\brX^{\brj}_{\mu}$ into (\ref{MilneG2}) reproduces (\ref{Bonly}). Alternatively, if we replace $V_{\mu i}$ and $\brV_{\mu\bri}$ in (\ref{MilneG2}) by $K_{\mu\nu}H^{\nu\rho}V_{\rho i}$ and $K_{\mu\nu}H^{\nu\rho}\brV_{\rho\bri}$, we recover (\ref{MilneG}).\\

The dynamics of the DFT-metric and the DFT-dilaton is dictated by the Euler-Lagrange equations of DFT. The expression of the  $(n,\brn)$ DFT-metric~(\ref{cHFINAL})  may then be inserted into the known stringy extension of the Christoffel symbol  to lead to covariant derivatives and curvatures~\cite{Jeon:2011cn}. Yet, the trace~(\ref{trace}) of the $(n,\brn)$ DFT-metric can be nontrivial, and  this calls for some revision of the previous result:  
\be
\ba{ll}
\Gamma_{CAB}=&2\left(P\partial_{C}P\brP\right)_{[AB]}
+2\left({{\brP}_{[A}{}^{D}{\brP}_{B]}{}^{E}}-{P_{[A}{}^{D}P_{B]}{}^{E}}\right)\partial_{D}P_{EC}\\
{}&-4\left(\textstyle{\frac{1}{P_{M}{}^{M}-1}}P_{C[A}P_{B]}{}^{D}+\textstyle{\frac{1}{\brP_{M}{}^{M}-1}}\brP_{C[A}\brP_{B]}{}^{D}\right)\!\left(\partial_{D}d+(P\partial^{E}P\brP)_{[ED]}\right)\,,
\ea
\label{Gammao}
\ee
which now allows for generic values for the traces  of the projectors,
\be
\ba{cc}
P_{M}{}^{M}=D+n-\brn\,,\quad&\quad
\brP_{M}{}^{M}=D-n+\brn\,.
\ea
\ee


\subsection{Particle and string on $(n,\brn)$ doubled-yet-gauged spacetime\label{SECsubPS}}

While the notion of  doubled-yet-gauged spacetime might sound  somewhat mysterious,    it is possible to define \textit{proper length}  and hence to show that it is a `metric space'.  To do so, we first   note that  the usual infinitesimal one-form, $\rd x^{A}$, is neither  diffeomorphism covariant~\eqref{diff1}, \eqref{diff2}, 
\be
\delta(\rd x^{A})=\rd x^{B}\partial_{B}V^{A}\neq\rd x^{B}(\partial_{B}V^{A}-\partial^{A}V_{B})\,,
\ee
 nor  coordinate gauge symmetry invariant~\eqref{CGS}, since 
\be
\rd\Delta^{A}=\rd x^{B}\partial_{B}\Delta^{A}\neq 0\,.
\ee 
Thus, the naive contraction with the DFT-metric, $\rd x^{A}\rd x^{B}\cH_{AB}$, cannot give  any sensible definition of   proper length in doubled-yet-gauged spacetime.   To cure the problem,  we  need to gauge $\rd x^{A}$  explicitly,   introducing  a connection, $\cA^{A}$,  which should satisfy the same property as the  coordinate gauge symmetry generator,  $\Delta^{A}$~\eqref{CGS},
\be
\ba{lll}
\rD x^{A}:=\rd x^{A}-\cA^{A}\,,\quad&\quad \cA^{A}\partial_{A}=0\,,\quad&\quad\cA^{A}\cA_{A}=0\,.
\ea
\label{DyGDx}
\ee
Provided the connection  transforms appropriately,  $\rD x^{A}$ becomes a well-behaved \textit{i.e.~}covariant    vector~\cite{Lee:2013hma},
\be
\ba{llll}
\delta x^{A}=\xi^{A}\,,\quad&\quad
\delta \cA^{A}=\partial^{A}\xi_{B}(\rd x^{B}-\cA^{B})\quad&\Longrightarrow&\quad
{\delta (\rD x^{A})=(\partial_{B}\xi^{A}-\partial^{A}\xi_{B})\rD x^{B}}\,;\\
\delta x^{A}=\Delta^{A}\,,\quad&\quad
\delta\cA^{A}=\rd\Delta^{A}\quad&\Longrightarrow&\quad{\delta(\rD x^{A})=0\,}.
\ea
\label{deltaU}
\ee
We propose then to define  the proper distance in doubled-yet-gauged spacetime  by  \textit{path integral}~\cite{JHPBanff},
\be
\left|\!\left| x_{1}\,,\, x_{2}\right|\!\right|:=-\ln\left[\int\cD\cA\,\exp\left(-\int_{1}^{2} \sqrt{\rD x^{A}\rD x^{B}\cH_{AB}}\right)\right]\,.
\label{Length}
\ee
By  letting    $\tpartial^{\mu}\equiv0$ and therefore $\cA^{A}\,{\equiv(A_{\mu},0)}$, we may solve  the  constraints    and write  
\be
\rD x^{A}\equiv\left(\rd\tx_{\mu}-A_{\mu}\,,\,\rd x^{\nu}\right)\,.
\ee 
That is to say,   only the half of the doubled coordinates, \textit{i.e.} $\tx_{\mu}$ directions,  are gauged.   Furthermore, with the Riemannian  DFT-metric~(\ref{RiemanncH}), we get~\cite{Lee:2013hma}
\be
\rD x^{A}\rD x^{B}\cH_{AB}
\equiv\rd{x}^{\mu}\rd{x}^{\nu}g_{\mu\nu}+
\left(\rd{\tx}_{\mu}-A_{\mu}+\rd{x}^{\rho}B_{\rho\mu}\right)
\left(\rd{\tx}_{\nu}-A_{\nu}+\rd{x}^{\sigma}B_{\sigma\nu}\right)
g^{\mu\nu}\,.
\label{RH}
\ee
Thus, after integrating out the auxiliary  connection, 
our proposal~\eqref{Length}  reduces -- at least classically --  to the conventional, \textit{i.e.~}Riemannian  proper distance,   $\left|\!\left| x_{1}^{\mu}\,,\, x_{2}^{\mu}\right|\!\right|={{\int_{1}^{2}}}\sqrt{\rd{x}^{\mu}\rd{x}^{\nu}g_{\mu\nu}}\,$. Being   independent of the    gauged  $\tx_{\mu}$ coordinates, \textit{i.e.}~$\left|\!\left| x_{1}^{A}\,,\, x_{2}^{A}\right|\!\right|\equiv\left|\!\left| x_{1}^{\mu}\,,\, x_{2}^{\mu}\right|\!\right|$,   indeed  the   formula~\eqref{Length} measures    the distance between two `gauge orbits'.\\

\noindent The exponent in \eqref{Length} immediately sets      the  action for a point particle in doubled-yet-gauged spacetime, or its  square-root free einbein formulation~\cite{Ko:2016dxa}, 
\be\dis{
S_{\scriptscriptstyle{\rm{particle}}}=
\int\rd\tau~e^{-1\,}\rD_{\tau}x^{A}\rD_{\tau}x^{B}\cH_{AB}(x)-\quarter m^{2}e\,.}
\label{particleaction}
\ee
It also   easily extends    to (Nambu-Goto type) area and volume, which in turn  provides the doubled-yet-gauged string action~\cite{Lee:2013hma} (\textit{c.f.~}\cite{Hull:2006va} and also \cite{Park:2016sbw}  for  the extension to the Green-Schwarz superstring),
\be
S_{\scriptscriptstyle{\rm{string}}}={\textstyle{\frac{1}{4\pi\alpha^{\prime}}}}{\dis{\int}}\rd^{2}\sigma~\cL_{\scriptscriptstyle{\rm{string}}}\,,
\qquad
\cL_{\scriptscriptstyle{\rm{string}}}=-\half\sqrt{-h}h^{\alpha\beta}\rD_{\alpha}x^{A}\rD_{\beta}x^{B}\cH_{AB}(x)-\epsilon^{\alpha\beta}\rD_{\alpha}x^{A}\cA_{\beta A}\,.
\label{stringaction}
\ee
These two actions are fully covariant under $\ODD$ rotations, coordinate gauge symmetry~(\ref{CGS}), target-spacetime  diffeomorphisms~(\ref{diff2}), world-volume diffeomorphisms and  Weyl symmetry in the string case.

Besides the constraint imposed by the auxiliary potential, $\cA^{A}$, the equation of motion of the former particle action can be spelled  in terms of the stringy Christoffel connection~(\ref{Gammao}),
\be
\textstyle{e\frac{\rd~}{\rd\tau}(e^{-1}\cH_{AB}\rD_{\tau}x^{B})+2\Gamma_{ABC}(\brP\rD_{\tau}x)^{B}(P\rD_{\tau}x)^{C}=0\,.}
\label{GEO0}
\ee
On the other hand,   for a string propagating on the $(0,0)$ Riemannian background, the auxiliary potential, $\cA^{A}$,  implies the self-duality (\textit{i.e.~}chirality)  over the entire doubled spacetime~\cite{Lee:2013hma}, 
\be
\rD_{\alpha}x_{A}+\textstyle{\frac{1}{\sqrt{-h}}}\epsilon_{\alpha}{}^{j\beta}\cH_{A}{}^{B}\rD_{\beta}x_{B}=0\,,
\ee
and the Euler-Lagrangian  equation of $x^{A}$ gets simplified to give   the \textit{stringy geodesic equation},
\be
\textstyle{\frac{1}{\sqrt{-h}}}\partial_{\alpha}(\sqrt{-h}\cH_{AB}\rD^{\alpha}x^{B})+\Gamma_{ABC}(\brP\rD_{\alpha}x)^{B}(P\rD^{\alpha}x)^{B}=0\,,
\label{GEO2}
\ee
which extends (\ref{GEO0}), yet with a different numerical factor in front of the  connection, $2$ \textit{versus} $1$.

For a generic  non-Riemannian background, the analysis is more subtle which we investigate hereafter.  We substitute  the generic $(n,\brn)$ DFT-metric~(\ref{cHFINAL}) into the covariant actions, and move from doubled formalism to undoubled one.   One useful identity which  generalizes (\ref{RH}) from  Riemannian $(0,0)$ to a generic  $(n,\brn)$ case is, with $\,\rD_{\alpha}x^{A}=(\partial_{\alpha}\tx_{\mu}-A_{\alpha\mu},\partial_{\alpha}x^{\nu})$, 
\be
\ba{lll}
\rD_{\alpha}x^{M}\rD_{\beta}x^{N}\cH_{MN}&=&\partial_{\alpha} x^{\mu}\partial_{\beta}x^{\nu}
K_{\mu\nu}+
\left(\rD_{\alpha}\tx_{\mu}-B_{\mu\kappa}\partial_{\alpha}x^{\kappa}\right)
\left(\rD_{\beta}\tx_{\nu}-B_{\nu\lambda}\partial_{\beta}x^{\lambda}\right)H^{\mu\nu}\\
{}&{}&~+2X^{i}_{\mu}\partial_{(\alpha}x^{\mu}\left[
\rD_{\beta)}\tx_{\nu}-B_{\nu\rho}\partial_{\beta)}x^{\rho}\right]Y_{i}^{\nu}
-2\brX^{\bri}_{\mu}\partial_{(\alpha}x^{\mu}\left[
\rD_{\beta)}\tx_{\nu}-B_{\nu\rho}\partial_{\beta)}x^{\rho}\right]\brY_{\bri}^{\nu}\,,
\ea
\label{useful0}
\ee
which reads more explicitly for particles, 
\be
\ba{lll}
\rD_{\tau}x^{M}\rD_{\tau}x^{N}\cH_{MN}&=
&\dot{x}^{\mu}\dot{x}^{\nu}
K_{\mu\nu}+
\left(\dot{\tx}_{\mu}-A_{\tau\mu}-B_{\mu\kappa}\dot{x}^{\kappa}\right)
\left(\dot{\tx}_{\nu}-A_{\tau\nu}-B_{\nu\lambda}\dot{x}^{\lambda}\right)H^{\mu\nu}\\
{}&{}&+2X^{i}_{\mu}\dot{x}^{\mu}\left(
\dot{\tx}_{\nu}-A_{\tau\nu}-B_{\nu\rho}\dot{x}^{\rho}\right)Y_{i}^{\nu}
-2\brX^{\bri}_{\mu}\dot{x}^{\mu}\left(
\dot{\tx}_{\nu}-A_{\tau\nu}-B_{\nu\rho}\dot{x}^{\rho}\right)\brY_{\bri}^{\nu}\,.
\ea
\label{DHD0}
\ee
Note that, in accordance with the completeness relation~(\ref{COMP}), the auxiliary vector potential decomposes as
\be
A_{\alpha\mu}=K_{\mu\nu}\left(H^{\nu\rho}A_{\alpha\rho}\right)
+X_{\mu}^{i}\left(Y^{\rho}_{i}A_{\alpha\rho}\right)
+\brX_{\mu}^{\bri}\left(\brY^{\rho}_{\bri}A_{\alpha\rho}\right)\,.
\ee

\begin{itemize}
\item[--] \textit{Particle dynamics.}\\
Integrating out $H^{\mu\nu}A_{\tau\nu}$  gives the on-shell relation,
\be
\ba{lll}
H^{\mu\nu}A_{\tau\nu}\equiv H^{\mu\nu}\left(\dot{\tx}_{\nu}-B_{\nu\lambda}\dot{x}^{\lambda}\right)
\quad&\quad\text{ or\, equivalently }\quad&\quad
H^{\mu\nu}\left(\rD_{\tau}\tx_{\nu}-B_{\nu\lambda}\dot{x}^{\lambda}\right)\equiv0\,,
\ea
\ee
which implies that the `dual' conjugate momenta  are trivial along   $D-n-\brn$ number of $\tx_{\mu}$ directions. 

On the other hand, integrating out the remaining components, $Y^{\rho}_{i}A_{\tau\rho}$ and $\brY^{\rho}_{\bri}A_{\tau\rho}$, we acquire  constraints on the $x^{\mu}$ coordinates,
\be
\ba{ll}
X^{i}_{\mu}\dot{x}^{\mu}\equiv0\,,\qquad&\qquad
\brX^{\bri}_{\mu}\dot{x}^{\mu}\equiv0\,.
\ea
\label{FROZEN}
\ee
Namely, the particle freezes over $n+\brn$ directions on the physical section formed by $x^{\mu}$ coordinates.

\item[--] \textit{String dynamics.}\\
For string, combining the useful identity~(\ref{useful0})  with the   topological term in the action~(\ref{stringaction}),  we can  reduce  the world-sheet  Lagrangian,
\be
\ba{lll}
\multicolumn{3}{c}{
\textstyle{\frac{1}{4\pi\alpha^{\prime}}}\cL_{\scriptscriptstyle{\rm{string}}}
=\textstyle{\frac{1}{2\pi\alpha^{\prime}}}
\cL^{\prime}_{\scriptscriptstyle{\rm{string}}}\,,}\\

\cL^{\prime}_{\scriptscriptstyle{\rm{string}}}&=&-\half\sqrt{-h}h^{\alpha\beta}\partial_{\alpha}x^{\mu}\partial_{\beta}x^{\nu}
K_{\mu\nu}
+\half\epsilon^{\alpha\beta}\partial_{\alpha}x^{\mu}
\partial_{\beta}x^{\nu}B_{\mu\nu}+\half\epsilon^{\alpha\beta}
\partial_{\alpha}\tx_{\mu}\partial_{\beta}x^{\mu}\\
{}&{}&-\half\sqrt{-h}h^{\alpha\gamma}\left[
X^{i}_{\mu}\left(\partial_{\alpha}x^{\mu}+\textstyle{\frac{1}{\sqrt{-h}}}\epsilon_{\alpha}{}^{\beta}\partial_{\beta}x^{\mu}\right)\right]
\left(\partial_{\gamma}\tx_{\nu}-A_{\gamma\nu}-B_{\nu\rho}\partial_{\gamma}x^{\rho}\right)Y_{i}^{\nu}\\
{}&{}&+\half\sqrt{-h}h^{\alpha\gamma}\left[
\brX^{\bri}_{\mu}\left(\partial_{\alpha}x^{\mu}-\textstyle{\frac{1}{\sqrt{-h}}}\epsilon_{\alpha}{}^{\beta}\partial_{\beta}x^{\mu}\right)\right]
\left(\partial_{\gamma}\tx_{\nu}-A_{\gamma\nu}-B_{\nu\rho}\partial_{\gamma}x^{\rho}\right)\brY_{\bri}^{\nu}\\
{}&{}&-\quarter\sqrt{-h}h^{\alpha\beta}\left(\cC_{\alpha\mu}-A_{\alpha\mu}\right)\left(\cC_{\beta\nu}-A_{\beta\nu}\right)H^{\mu\nu}\,,
\ea
\label{Lpr}
\ee
where for shorthand notation we set
\be
\cC_{\alpha\mu}:=\partial_{\alpha}\tx_{\mu}-B_{\mu\nu}\partial_{\alpha}x^{\nu}+\textstyle{\frac{1}{\sqrt{-h}}}\epsilon_{\alpha}{}^{\beta}K_{\mu\nu}\partial_{\beta}x^{\nu}\,.
\ee
Now,  integrating out  $H^{\mu\nu}A_{\alpha\nu}$ we obtain the on-shell relation,
\be
\ba{lll}
H^{\mu\nu}A_{\alpha\nu}\equiv H^{\mu\nu}\cC_{\alpha\nu}\quad&~~\text{ or\, equivalently }~~&\quad
H^{\mu\nu}\left(\rD_{\alpha}\tx_{\mu}-B_{\mu\nu}\partial_{\alpha}x^{\nu}+\textstyle{\frac{1}{\sqrt{-h}}}\epsilon_{\alpha}{}^{\beta}K_{\mu\nu}\partial_{\beta}x^{\nu}\right)\equiv 0\,,
\ea
\ee
and  integrating out  $Y^{\nu}_{i}A_{\alpha\nu}$,  $\brY^{\nu}_{\bri}A_{\alpha\nu}$,  we obtain chiral constraints,
\be
\ba{ll}
X^{i}_{\mu}\left(\partial_{\alpha}x^{\mu}+\textstyle{\frac{1}{\sqrt{-h}}}\epsilon_{\alpha}{}^{\beta}\partial_{\beta}x^{\mu}\right)\equiv0\,,
\quad&\quad
\brX^{\bri}_{\mu}\left(\partial_{\alpha}x^{\mu}-\textstyle{\frac{1}{\sqrt{-h}}}\epsilon_{\alpha}{}^{\beta}\partial_{\beta}x^{\mu}\right)\equiv0\,.
\ea
\ee
Namely, string becomes chiral over $n$ directions and anti-chiral over $\brn$ directions on the section coordinatized by $x^{\mu}$. The chirality further implies that  strings which meet  boundary conditions (periodic, Neumann or Dirichlet) are also frozen,  or localized,  over the $n+\brn$ directions, similarly to the particle case~(\ref{FROZEN}).

\end{itemize}


\subsection{DFT-vielbeins for $(n,\brn)$ doubled-yet-gauged spacetime\label{SECvielbein}}
In order to  couple to  fermions~\cite{Jeon:2011vx,Choi:2015bga} or R-R sector~\cite{Jeon:2012kd}, as well as for supersymmetrizations~\cite{Jeon:2011sq,Jeon:2012hp,Cho:2015lha},  it is necessary to introduce   \textit{a pair of} DFT-vielbeins,   $V_{Ap}$ and $\brV_{A\brp}$,  from which one can construct the projectors,
\be
\ba{ll}
P_{AB}=\half(\cJ_{AB}+\cH_{AB})=V_{Ap}V_{Bq}\eta^{pq}\,,\quad&\quad
\brP_{AB}=\half(\cJ_{AB}-\cH_{AB})=\brV_{A\brp}\brV_{B\brq}\breta^{\brp\brq}\,,
\ea
\ee 
where   $\eta_{pq}$ and $\breta_{\brp\brq}$ are the two  constant metrics of   twofold local Lorentz symmetries for   two distinct   locally inertial frames, one for the left and the other  for the right closed string modes~\cite{Duff:1986ne}.\footnote{Ensuring the twofold  spin  groups in the maximally supersymmetric  DFT~\cite{Jeon:2012hp} and the doubled-yet-gauged  Green-Schwarz superstring action~\cite{Park:2016sbw}, the conventional IIA and IIB theories are unified into a single  theory that is chiral with respect to both spin groups.  The distinction of  IIA and IIB then refers to   two different types of  (Riemannian)  `solutions' rather than `theories'.} To ensure the symmetric, orthogonal and completeness properties of the projectors~(\ref{Pdef}), the DFT-vielbeins must satisfy their own defining properties:    
\be
\ba{llll}
V_{Ap}V^{A}{}_{q}=\eta_{pq}\,,\quad&\quad 
\brV_{A\brp}\brV^{A}{}_{\brq}=\breta_{\brp\brq}\,,\quad&\quad V_{Ap}\brV^{A\brp}=0\,,\quad&\quad 
V_{Ap}V_{B}{}^{p}+\brV_{A\brp}\brV_{B}{}^{\brp}=\cJ_{AB}\,.
\ea
\label{defV}
\ee
Essentially, with  $\cH_{AB}=V_{Ap}V_{B}{}^{p}-\brV_{A\brp}\brV_{B}{}^{\brp}$,   DFT-vielbeins  diagonalize $\cJ_{AB}$ and $\cH_{AB}$ simultaneously, with the eigenvalues, $(\eta,+\breta)$ and $(\eta,-\breta)$.

\noindent The main result of this subsection  is the construction of the DFT-vielbeins, $V_{Ap}$ and $\brV_{A\brp}$,  for the general $(n,\brn)$    DFT-metric~(\ref{cHFINAL}). They are given by  $2D\times (D+n-\brn)$ and $2D\times (D-n+\brn)$ matrices respectively,
\be
\ba{ll}
V_{Ap}=\frac{1}{\sqrt{2}}\left(\ba{c}\cY_{p}{}^{\mu}\\\cX_{\nu}{}^{q}\eta_{qp}+B_{\nu\sigma}\cY_{p}{}^{\sigma}
\ea\right)\,,\quad&\quad
\brV_{A\brp}=\frac{1}{\sqrt{2}}\left(\ba{c}\brcY_{\brp}{}^{\mu}\\\brcX_{\nu}{}^{\brq}\breta_{\brq\brp}+B_{\nu\sigma}\brcY_{\brp}{}^{\sigma}
\ea\right)\,.
\ea
\label{VbrV}
\ee
Here $\cX_{\mu}{}^{p}$, $\cY_{p}{}^{\mu}$, $\brcX_{\nu}{}^{\brq}$ and $\brcY_{\brq}{}^{\nu}$ are respectively  $D\times(D+n-\brn)$, $\,(D+n-\brn)\times D$, $\,D\times(D-n+\brn)$ and $\,(D-n+\brn)\times D$  matrices,  such that $1\leq p\leq D+n-\brn$ \,and\, $1\leq\brp\leq D-n+\brn$.   Explicitly, with the smaller range of indices, $1\leq a,\bra \leq D{-n-\brn}$ and  $1\leq i\leq n$, $~1\leq\bri\leq\brn$ as before,   the matrices  read
\be
\ba{lll}
\ba{l}
\cX_{\mu}{}^{p}:=\left(\ba{lll}k_{\mu}{}^{a}& X_{\mu}^{i}& X_{\mu}^{j}\ea\right)\,,\\
\brcX_{\mu}{}^{\brp}:=\left(\ba{lll}\brk_{\mu}{}^{\bra}&\,\brX_{\mu}^{\bri}& \,\brX_{\mu}^{\brj}\ea\right)\,,
\ea\quad&\quad
\cY_{p}{}^{\mu}:=\left(\ba{c}h_{a}{}^{\mu}\\Y^{\mu}_{i}\\Y^{\mu}_{j}\ea\right)\,,\quad&\quad
\brcY_{\brp}{}^{\mu}:=\left(\ba{c}\brh_{\bra}{}^{\mu}\\\brY^{\mu}_{\bri}\\\brY^{\mu}_{\brj}\ea\right)\,,
\ea
\label{cXcY}
\ee 
where $\{h_{a}{}^{\mu},k_{\nu}{}^{b}\}$ and $\{\brh_{\bra}{}^{\mu},\brk_{\nu}{}^{\brb}\}$ are two  sets of  the ``square-roots'' of $H^{\mu\nu}$ and $K_{\mu\nu}$,
\be
\ba{ll}
H^{\mu\nu}=\eta^{ab}h_{a}{}^{\mu}h_{b}{}^{\nu}=-\breta^{\bra\brb}\brh_{\bra}{}^{\mu}\brh_{\brb}{}^{\nu}\,,\quad&\quad
K_{\mu\nu}=k_{\mu}{}^{a}k_{\nu}{}^{b}\eta_{ab}=-\brk_{\mu}{}^{\bra}\brk_{\nu}{}^{\brb}\breta_{\bra\brb}\,.
\ea
\label{HKhk}
\ee
The `total' twofold  local Lorentz symmetry group is clearly  $\Spin(t+n,s+n)\times\Spin(s+\brn,t+\brn)$,  with  $t+s+n+\brn=D$, where $(t,s)$ is the signature of $H^{\mu\nu}$ and $K_{\mu\nu}$.  The corresponding constant metrics are $\eta_{pq}$ and $\breta_{\brp\brq}$ respectively, while   $\eta_{ab}$ and $\breta_{\bra\brb}$ are  $(t+s)\times(t+s)$ sub-blocks of them,   of which the signatures  are  numerically opposite to each other~\cite{Jeon:2011cn},
\be
\ba{ll}
\eta_{pq}=\left(\ba{ccc}\eta_{ab}&{0}&{0}\\{0}&-\delta_{ij}&{0}\\
{0}&{0}&+\delta_{ij}\ea\right)\,,\quad&\quad
\eta_{ab}=\mbox{diag}(\underbrace{-\,-\,\cdots\, -\,-}_{t}\,\underbrace{+\,+\,\cdots \,+\,+}_{s})\,,\\
\breta_{\brp\brq}=\left(\ba{ccc}
\breta_{\bra\brb}&{0}&{0}\\{0}&+\delta_{\bri\brj}&{0}\\
{0}&{0}&-\delta_{\bri\brj}\ea\right)\,,
\quad&\quad
\breta_{\bra\brb}=\mbox{diag}(\underbrace{+\,+\,\cdots\, +\,+}_{t}\,\underbrace{-\,-\,\cdots \,-\,-}_{s})\,.
\ea
\label{etas}
\ee

\noindent There are defining properties of $\{h_{a}{}^{\mu},k_{\nu}{}^{b}\}$ and $\{\brh_{\bra}{}^{\mu},\brk_{\nu}{}^{\brb}\}$, in accordance with \eqref{HXX}, \eqref{COMP}: 
\be
\ba{llll}
h_{a}{}^{\mu}X_{\mu}^{i}=0\,,\quad&\quad
h_{a}{}^{\mu}\brX_{\mu}^{\bri}=0\,,\quad&\quad
Y_{i}^{\mu}k_{\mu}{}^{a}=0\,,\quad&\quad
\brY_{\bri}^{\mu}k_{\mu}{}^{a}=0\,,\\
\brh_{\bra}{}^{\mu}X_{\mu}^{i}=0\,,\quad&\quad
\brh_{\bra}{}^{\mu}\brX_{\mu}^{\bri}=0\,,\quad&\quad
Y_{i}^{\mu}\brk_{\mu}{}^{\bra}=0\,,\quad&\quad
\brY_{\bri}^{\mu}\brk_{\mu}{}^{\bra}=0\,,\\
\multicolumn{4}{c}{
k_{\mu}{}^{a}h_{a}{}^{\nu}+X_{\mu}^{i}Y_{i}^{\nu}
+\brX_{\mu}^{\bri}\brY_{\bri}^{\nu}=\delta_{\mu}{}^{\nu}\,,\quad\quad
h_{a}{}^{\mu}k_{\mu}{}^{b}=\delta_{a}{}^{b}\,,}\\
\multicolumn{4}{c}{
\brk_{\mu}{}^{\bra}\brh_{\bra}{}^{\nu}+X_{\mu}^{i}Y_{i}^{\nu}
+\brX_{\mu}^{\bri}\brY_{\bri}^{\nu}=\delta_{\mu}{}^{\nu}\,,\quad\quad
\brh_{\bra}{}^{\mu}\brk_{\mu}{}^{\brb}=\delta_{\bra}{}^{\brb}\,.}
\ea
\label{COMP2}
\ee
It follows that 
\be
\ba{ll}
\cX_{\mu}{}^{p}\cY_{p}{}^{\nu}=
\delta_{\mu}{}^{\nu}+X_{\mu}^{i}Y_{i}^{\nu}-\brX_{\mu}^{\bri}\brY_{\bri}^{\nu}
\,,\quad&\quad 
\brcX_{\mu}{}^{\brp}\brcY_{\brp}{}^{\nu}=\delta_{\mu}{}^{\nu}-X_{\mu}^{i}Y_{i}^{\nu}+\brX_{\mu}^{\bri}\brY_{\bri}^{\nu}\,,\\
\vspace{12pt}
\cY_{p}{}^{\lambda}\cX_{\lambda}{}^{q}=
\left(\ba{ccc}
\delta_{a}{}^{b}&0&0\\
0&\delta_{i}{}^{k}&\delta_{i}{}^{l}\\
0&\delta_{j}{}^{k}&\delta_{j}{}^{l}
\ea\right)\,,
\quad&\quad
\brcY_{\brp}{}^{\lambda}\brcX_{\lambda}{}^{\brq}=
\left(\ba{ccc}
\delta_{\bra}{}^{\brb}&0&0\\
0&\delta_{\bri}{}^{\brk}&\delta_{\bri}{}^{\brl}\\
0&\delta_{\brj}{}^{\brk}&\delta_{\brj}{}^{\brl}
\ea\right)\,,
\ea
\ee
and
\be
\ba{lll}
P_{AB}&=&\left(\ba{cc}
\half H&Y_{i}(X^{i})^{T}+\half H(K-B)\\
X^{i}(Y_{i})^{T}+\half (K+B)H\quad&\quad \half(K+B)H(K-B)+BY_{i}(X^{i})^{T}-X^{i}(Y_{i})^{T}B\ea\right)\,,\\
\brP_{AB}&=&\left(\ba{cc}
-\half H&\brY_{\bri}(\brX^{\bri})^{T}+\half H(K+B)\\
\brX^{\bri}(\brY_{\bri})^{T}+\half (K-B)H\quad&\quad -\half(K-B)H(K+B)+B\brY_{\bri}(\brX^{\bri})^{T}-\brX^{\bri}(\brY_{\bri})^{T}B\ea\right)\,,
\ea
\ee
where the superscript $\scriptstyle{T}$ converts column vectors to row ones. As expected, $P_{AB}$ and $\brP_{AB}$ are respectively   free of the barred  and unbarred variables, $\{\brX^{\bri},\brY_{\brj}\}$ and $\{X^{i},Y_{j}\}$.

\newpage
\noindent In a parallel manner to (\ref{Bcontri}), the $B$-field contributes to the DFT--vielbeins through $\ODD$ multiplications, 
\be
\ba{ll}
V_{Mp}=\frac{1}{\sqrt{2}}\left(\ba{cc}1&0\\B&1\ea\right)\left(\ba{c}\cY^{T}\\\cX\eta
\ea\right)\,,\quad&\quad
\brV_{M\brp}=\frac{1}{\sqrt{2}}\left(\ba{cc}1&0\\B&1\ea\right)\left(\ba{c}\brcY^{T}\\\brcX\breta
\ea\right)\,.
\ea
\label{BcontriV}
\ee
For consistency, the trace of the DFT-metric reads
\be
\cH_{A}{}^{A}=\eta_{p}{}^{p}-\breta_{\brp}{}^{\brp}=(t+s+2n)-(t+s+2\brn)=(D+n-\brn)-(D-n+\brn)=2(n-\brn)\,.
\label{cHtrace}
\ee

\noindent The symmetry of the DFT-metric~(\ref{MilneG2}) extends to  DFT-vielbeins:
\be
\ba{lll}
Y_{i}^{\mu}\quad&\longmapsto&\quad Y_{i}^{\mu}+H^{\mu\nu}V_{\nu i}\,,\\
\brY_{\bri}^{\mu}\quad&\longmapsto&\quad\brY_{\bri}^{\mu}+H^{\mu\nu}\brV_{\nu\bri}\,,\\
k_{\mu}{}^{a}\quad&\longmapsto&\quad k_{\mu}{}^{a}-X_{\mu}^{i}\eta^{ab}h_{b}{}^{\nu}V_{\nu i}
-\brX_{\mu}^{\bri}\eta^{ab}h_{b}{}^{\nu}\brV_{\nu\bri}\,,\\
\brk_{\mu}{}^{\bra}\quad&\longmapsto&\quad\brk_{\mu}{}^{\bra}+X_{\mu}^{i}\breta^{\bra\brb}\brh_{\brb}{}^{\nu}V_{\nu i}+\brX_{\mu}^{\bri}\breta^{\bra\brb}\brh_{\brb}{}^{\nu}\brV_{\nu\bri}\,,\\
B_{\mu\nu}\quad&\longmapsto&\quad B_{\mu\nu}
-2X^{i}_{[\mu}V_{\nu]i}+2\brX^{\bri}_{[\mu}\brV_{\nu]\bri}
+2X^{i}_{[\mu}\brX^{\bri}_{\nu]}\left(Y_{i}^{\rho}\brV_{\rho\bri}
+\brY_{\bri}^{\rho}V_{\rho i}+V_{\rho i}H^{\rho\sigma}\brV_{\sigma\bri}\right)\,.
\ea
\label{MilneG3}
\ee
Under these transformations, the DFT-vielbeins are invariant up to twofold local Lorentz rotations. 

\noindent As seen from the  doubled-yet-gauged actions for  particle and string~(\ref{DHD0}), (\ref{Lpr}), as well as the  coupling to a scalar field~(\ref{Scalar}),   it is not the full signatures of the spin group, \bea\Spin(t+n,s+n)\times\Spin(s+\brn,t+\brn)\, ,\eea but  the signature of $K_{\mu\nu}$ and $H^{\mu\nu}$, \textit{i.e.~}$(t,s)$, that matters for   \textit{unitarity.}   The choice of $t=1$ then amounts to the usual  Minkowskian spacetime.


\subsection{Kaluza-Klein ansatz for DFT}
The ordinary Kaluza-Klein  ansatz for a Riemannian metric can be `block-diagonalized',
\be
\small{
\ba{lll}
\hat{g}={{\left(\ba{cc}g^{\prime}+aga^{T} & ag\\ ga^{T}& g\ea
\right)}}=\exp\left[\hat{a}\right]{{\left(\ba{cc} g^{\prime}&0\\ 0& g\ea\right)}}\exp\left[\hat{a}^{T}\right]
~~~~&\mbox{~~where~~}&~~~~ \hat{a}_{\hat{\mu}}{}^{\hat{\nu}}={{\left(\ba{cc} 0&a_{\mu^{\prime}}{}^{\nu}\\0&0\ea\right)}}\,.
\ea}
\label{ordinaryKK}
\ee
In a similar fashion, we propose the  Kaluza-Klein ansatz for the DFT-metric, $\hcH_{\hM\hN}$, 
\be
\hcH=\exp\left[\hW\,\right]\left(\ba{cc} 
\cHpr &0 \\ 0& \cH\ea\right)\exp\left[\hW^{T}\,\right]\,,
\ee
for which we decompose $\hD=D^{\prime}+D$,  such that
\be
\ba{ll}
\hODD ~~\rightarrow~~ \ODDpr \times \ODD\,,\quad&\quad
\hcJ=\left(\ba{cc}\cJ^{\prime}&0\\0&\cJ\ea\right)\,,
\ea
\ee
and set an off-block-diagonal  $\mathfrak{so}(D,D)$ element,
\be
\ba{ll}
\hW= \left(\ba{cc} 0&-W\\\brW&0\ea\right) \in \mathfrak{so}(D,D)\,,\quad&\quad
\brW_{M}{}^{\Mpr}:=W^{\Mpr}{}_{M}= \cJ_{MN}W_{\Npr}{}^{N}\cJ^{\prime\Npr\Mpr}\,.
\ea
\ee
Further, we impose a  constraint on the $2\Dpr\times 2D$ matrix,  $W_{\Mpr}{}^{N}$, 
\be
\ba{lll}
\brW W=0\quad&~~\text{ or\, explicitly }~~&\quad W_{\Lpr M}W^{\Lpr N}=0\,,
\ea
\label{Wconstraint} 
\ee
which  sets half of its components trivial.  At least for the Riemannian, \textit{i.e.~}$(0,0)$ case, this constraint  makes the counting of the degrees of freedom consistent: $g_{\mu\nu}$ and $B_{\mu\nu}$ have  $D^{2}$ degrees of freedom, while  $W_{\Mpr}{}^{N}$ has $2\Dpr D$ degrees, such that
\be
\hat{D}^{2}=(\Dpr+D)^{2}=\Dpr{}^{2}+D^{2}+2\Dpr D\,,
\ee
matching the degrees of freedom between $\hat{\cH}$ and $\{\cH^{\prime},\cH,W\}$.  Essentially, $\hat{g}_{\mu^{\prime}\nu}$ and $\hat{B}_{\mu^{\prime}\nu}$ constitute $W_{M^{\prime}}{}^{N}$.

Explicitly, we have $\hW^{3}=0$ and 
\be
\hcH=\left(\ba{cc}
(1-\half W\brW)\cH^{\prime}(1-\half W\brW)^{T}+W\cH W^{T}~~~~&~~~~-W\cH+(1-\half W\brW)\cHpr\brW^{T}\\
-\cH W^{T}+\brW\cH^{\prime}(1-\half W\brW)^{T}~~~~&~~~~\cH+\brW\cHpr\brW^{T}\ea\right)\,,
\ee
which is classified by four non-negative integers: $(n,\brn)$ for $\cH_{AB}$ and  $(n^{\prime},\brn^{\prime})$ for $\cH^{\prime}_{\Apr\Bpr}$, with the total trace, $\hcH_{\hA}{}^{\hA}=2(n+n^{\prime}-\brn-\brn^{\prime})$.

Especially,   in the maximally non-Riemannian case of $\cH^{\prime}=\cJ^{\prime}$,  \textit{i.e.~}$(n^{\prime},\brn^{\prime})=(D^{\prime},0)$, the above expression dramatically  simplifies
\be
\hcH=\left(\ba{cc}
\cJ^{\prime}-2W\brP W^{T}\,&\,2W\brP\\
2\brP W^{T}\,&\,\cH\ea\right)\,.
\label{dramatic}
\ee
~\\
\noindent Intriguingly, the resulting field content, $\cH_{AB}, \brP_{AB}W_{A^{\prime}}{}^{B}$,  coincides with the ansatz for  heterotic DFT proposed by Hohm, Sen and Zwiebach~\cite{Hohm:2014sxa}. We leave it as a future work to explore the tantalizing connection between  heterotic string and  non-Riemannian doubled-yet-gauged spacetime, possibly using  the Scherk-Schwartz reduction scheme in DFT~\cite{Geissbuhler:2011mx,Aldazabal:2011nj,
Grana:2012rr,Geissbuhler:2013uka,Berman:2013cli,Cho:2015lha,Malek:2016vsh,Malek:2017njj,PREP}.

\section{Applications\label{SECA}}
The case of $(0,0)$ admits a  well-defined Riemannian metric and hence corresponds to Riemannian geometry, or to ``Generalized Geometry''~\cite{Hitchin:2004ut,Gualtieri:2003dx,
Hitchin:2010qz,Coimbra:2011nw,Coimbra:2012yy,Garcia-Fernandez:2013gja} when equipped with the pair of  DFT-vielbeins.  In this section,   
we discuss various  applications of other  $(n,\brn)$ backgrounds and identify the corresponding geometries. 

\subsection{Maximally non-Riemannian $(D,0)$\,: Siegel's chiral string}
In the  maximally non-Riemannian case of $(D,0)$, with $i=1,2,\cdots, D$,  we can view $X^{i}_{\mu}$  as a non-degenerate $D\times D$ square matrix. Then from (\ref{complete}) and 
\be
\left(X^{j}_{\lambda}Y_{j}^{\mu}\right)X^{i}_{\mu}
=X^{i}_{\lambda}\,,
\ee
we conclude that $X^{j}_{\lambda}Y_{j}^{\mu}$ is actually an identity,
\be
X^{j}_{\lambda}Y_{j}^{\mu}=\delta_{\lambda}{}^{\mu}\,.
\ee
Thus,  in the case of $(D,0)$, we have
\be
\ba{ll}
\cJ_{AB}=\cH_{AB}=P_{AB}\,,\qquad&\qquad \brP_{AB}=0\,.
\ea
\ee
The corresponding  DFT-vielbein, $V_{Ap}$~(\ref{VbrV}) and the $\Spin(D,D)$ metric are   also $2D\times 2D$ square matrices,
\be
\ba{ll}
V_{Ap}=\textstyle{\frac{1}{\sqrt{2}}}\left(\ba{rr}
1&1\\
-1&1
\ea\right)\,,\qquad&\qquad
\eta_{pq}=\left(\ba{cc}-\delta_{ij}&~~0\\
~~0&~~+\delta_{ij}\ea\right)\,.
\ea
\ee
On the other hand, $\brV_{A\brp}$ is trivial.

The resulting  string action  is completely chiral on the $D$-dimensional section~(\ref{Lpr})~\cite{Lee:2013hma},
\be
\ba{ll}
S_{\scriptscriptstyle{\rm{string}}}={\textstyle{\frac{1}{4\pi\alpha^{\prime}}}}{\dis{\int}}\rd^{2}\sigma~~\epsilon^{\alpha\beta}
\partial_{\alpha}\tx_{\mu}\partial_{\beta}x^{\mu}\,,\qquad&\qquad
\partial_{\alpha}x^{\mu}+\textstyle{\frac{1}{\sqrt{-h}}}\epsilon_{\alpha}{}^{\beta}\partial_{\beta}x^{\mu}=0\,.
\ea
\label{chiralstring}
\ee
From the conventional $(0,0)$ set-up,  noting the sign difference,
\be
\ba{ll}
\cJ_{AB}=V_{A}{}^{p}V_{B}{}^{q}\eta_{pq}
+\brV_{A}{}^{\brp}\brV_{B}{}^{\brq}\breta_{\brp\brq}\,,\qquad&\qquad
\cH_{AB}=V_{A}{}^{p}V_{B}{}^{q}\eta_{pq}
-\brV_{A}{}^{\brp}\brV_{B}{}^{\brq}\breta_{\brp\brq}\,,
\ea
\ee
we may regard the substitution  of the $\ODD$ invariant metric, $\cJ_{AB}$, into  the DFT-metric, $\cH_{AB}$, inside the doubled-yet-gauged string action~(\ref{stringaction}) as the flipping of the  spin group signature, 
\be
\ba{lll}
\breta_{\brp\brq}\quad&\quad\longrightarrow\quad&\quad
- \breta_{\brp\brq}\,,
\ea
\ee 
such that $\eta_{pq}$ and $-\breta_{\brp\brq}$ assume  not opposite~\eqref{etas} but rather identical signatures. That is to say, there are no right modes: only left modes exist. This is consistent with  \eqref{chiralstring}, and  realizes the chiral string theory \textit{\`a la} Siegel~\cite{Siegel:2015axg}\footnote{See also \cite{Hohm:2013jaa,Huang:2016bdd}.} in a  rather  geometric  set-up.

  \subsection{$D=10,$ $(3,3)$\,: Non-Riemannian dimensional  reduction from ten to four}
 If we set $n=\brn$, then the DFT-metric is traceless and the two spin groups become commonly  $D$-dimensional,
\be
\ba{lll}
\Spin(t+n,s+n)\times\Spin(s+n,t+n)
\quad&\quad\mbox{where}~~&\quad t+s+2n=D\,.
\ea
\ee 
Thus,  the maximally supersymmetric $D=10$ DFT~\cite{Jeon:2012hp} and  the doubled-yet-gauged Green-Schwarz superstring~\cite{Park:2016sbw}, both of which assume    the Minkowskian Spin group,    
$\Spin(1,9)\times\Spin(9,1)$,  can accommodate  $(0,0)$ and $(1,1)$.   However, the theories constructed in \cite{Jeon:2012hp,Park:2016sbw}  can be readily  generalized  to  an arbitrary  signature, $\Spin(\hat{t},\hat{s})\times\Spin(\hat{s},\hat{t})$, with $\hat{t}+\hat{s}=10$,  by relaxing the Majorana condition on the  spinors    and  employing  their charge conjugations only, without involving the complex Dirac conjugations.  In this  case, the theory can describe $(n,n)$ non-Riemannian doubled-yet-gauged spacetime with  $n=0,1,2,\cdots,\min\!\left(\hat{t},\hat{s}\right)$.

An interesting choice then appears to be  $\Spin(4,6)\times\Spin(6,4)$.  Such a choice can encompass    six-dimensional  $(3,3)$  non-Riemannian   `internal' spacetime,  while maintaining  the ordinary    four-dimensional   Minkowskian   `external' spacetime.  As analyzed in subsection~\ref{SECsubPS},  point particles and  strings   freeze  on the $(3,3)$ internal spacetime and this may imply a natural  dimensional reduction  of string theory   from ten to four, alternative to the conventional   compactification on `small' Riemannian manifolds, \textit{e.g.~}$\mbox{CY}_{3}$. The latter will be of interest to analyze the Killing spinor equations in ~\cite{Jeon:2012hp} for the $D=10$ $(3,3)$ DFT-vielbeins~(\ref{VbrV}).  Certainly, constant  `flat' backgrounds are maximally  supersymmetric.


\subsection{$(1,1)$\,: Non-relativistic limit \textit{\`a la}~Gomis-Ooguri\label{SEC1c1}}

In this subsection, we identify  $(1,1)$  as the non-relativistic limit \textit{\`a la}~Gomis-Ooguri~\cite{Gomis:2000bd}. We start by considering a \textit{generic} Riemannian metric which depends  explicitly on the speed of light, $c$, 
\be
g_{\mu\nu}=-c^{2}T_{\mu}T_{\nu}(1-S^{\rho}S^{\sigma} \Phi_{\rho\sigma})+2cT_{(\mu}\Phi_{\nu)\rho}S^{\rho}+\Phi_{\mu\nu}\,,
\label{gc}
\ee
where $T_{\mu}$ and $S^{\nu}$ are orthogonal  time-like  and space-like vectors,
\be
T_{\mu}S^{\mu}=0\,.
\label{TS}
\ee
Essentially, (\ref{gc}) is  the `covariantized' form of the ordinary Kaluza-Klein ansatz for the Riemannian metric~(\ref{ordinaryKK}) as
\be
\ba{ll}
g_{\mu\nu}=(\delta_{\mu}{}^{\rho}+cT_{\mu}S^{\rho})
(\delta_{\nu}{}^{\sigma}+cT_{\nu}S^{\sigma})
(-c^{2}T_{\rho}T_{\sigma}+\Phi_{\rho\sigma})\,,\quad&\quad
\left[\exp(cT{\cdot S})\right]_{\mu}{}^{\nu}=\delta_{\mu}{}^{\nu}+cT_{\mu}S^{\nu}\,.
\ea
\ee
The inverse of the metric is then given by
\be
g^{\mu\nu}
=\Upsilon^{\mu\nu}-S^{\mu}S^{\nu}+\textstyle{\frac{2}{c}}N^{(\mu}S^{\nu)}-\textstyle{\frac{1}{c^{2}}}N^{\mu}N^{\nu}=(\delta^{\mu}{}_{\rho}-cS^{\mu}T_{\rho})
(\delta^{\nu}{}_{\sigma}-cS^{\nu}T_{\sigma})
(-\textstyle{\frac{1}{c^{2}}}N^{\rho}N^{\sigma}+\Upsilon^{\rho\sigma})\,,
\ee
where the variables, $N^{\nu}$ and $\Upsilon^{\mu\nu}$, meet by construction,\footnote{In subsections~\ref{SECDm1} and \ref{SEC10}, $\{T_{\mu}, N^{\nu}, \Upsilon^{\mu\nu}, \Phi_{\mu\nu}\}$  will be identified as either Carroll or  Newton-Cartan variables.} 
\be
\ba{llll}
T_{\mu}N^{\mu}=1\,,\quad&\quad
T_{\mu}\Upsilon^{\mu\nu}=0\,,\quad&\quad
N^{\mu}\Phi_{\mu\nu}=0\,,\quad&\quad
T_{\mu}N^{\nu}+\Phi_{\mu\rho}\Upsilon^{\rho\nu}=\delta_{\mu}{}^{\nu}\,.
\ea
\label{TNUpsilon}
\ee
Now, we  introduce an  ansatz for the $B$-field in a similar manner, 
\be
B_{\mu\nu}=2cT_{[\mu}B_{\nu]}+B_{\mu\nu}^{\scriptscriptstyle{0}}\,,
\label{Bansatz}
\ee
and require  that the Riemannian DFT-metric~(\ref{RiemanncH})  should be   non-singular in the non-relativistic,    large $c$ limit. In \eqref{Bansatz},  without loss of generality we may set $B_{\mu}$ to be orthogonal to $N^{\nu}$, \textit{i.e.~}$N^{\mu}B_{\mu}=0$. Further,  $B_{\mu\nu}^{\scriptscriptstyle{0}}$ denotes the  zeroth order in $c$ which is arbitrary and should survive once the limit is taken, as expected from the `Abelian' nature of the $B$-field from \eqref{Bcontri} and  \eqref{BAbelian}.

Clearly in the limit of $c\rightarrow\infty$, the inverse,  $g^{\mu\nu}$, is regular. We only need to ensure  both  $g^{-1}B$ and $g-Bg^{-1}B$ to be non-singular. The former implies
\be
\ba{ll}
\left(\Upsilon^{\mu\nu}-S^{\mu}S^{\nu}\right)B_{\mu}=0\,,\quad&\quad
\dis{\lim_{c\rightarrow\infty}}
g^{\mu\rho}B_{\rho\nu}=(\Upsilon^{\mu\rho}-S^{\mu}S^{\rho})B_{\rho\nu}^{\scriptscriptstyle{0}}+S^{\mu}B_{\nu}-(S^{\rho}B_{\rho})N^{\mu}T_{\nu}\,.
\ea
\label{USSB}
\ee
In turn, $Bg^{-1}B$ cannot be quadratically singular, and hence for the regularity of  $g-Bg^{-1}B$, the leading power of $g$  must be first order in $c$, \ie the apparent second order term in \eqref{gc} must be trivial, 
\be
S^{\rho}S^{\sigma}\Phi_{\rho\sigma}=1\,.
\label{S2}
\ee
Therefore,  the nontrivial cancellation of diverging terms inside $g-Bg^{-1}B$ takes place  at the first order, which reads
\be
c\times\Big[\left(\Phi_{\mu\rho}S^{\rho}-B_{\rho}S^{\rho}B_{\mu}\right)T_{\nu}+\left(\Phi_{\nu\rho}S^{\rho}-B_{\rho}S^{\rho}B_{\nu}\right)T_{\mu}\,\Big]=0\,.
\ee
Contraction of the quantity inside the square bracket  with $N^{\nu}$  gives  
\be
B_{\rho}S^{\rho}B_{\mu}=\Phi_{\mu\rho}S^{\rho}\,.\label{BSB}
\ee
Hence from (\ref{S2}) and (\ref{BSB}), we obtain
\be
B_{\rho}S^{\rho}=\pm 1\,.
\ee 
It follows that $g-Bg^{-1}B$ is non-singular as
\be
\ba{lll}
\dis{\lim_{c\rightarrow\infty}}\left(
g_{\mu\nu}-B_{\mu\rho}g^{\rho\sigma}B_{\sigma\nu}\right)&
=&\Phi_{\mu\nu}-B_{\mu}B_{\nu}-
B^{\scriptscriptstyle{0}}_{\mu\rho}(\Upsilon^{\rho\sigma}-S^{\rho}S^{\sigma})B_{\sigma\nu}^{\scriptscriptstyle{0}}
\\
{}&{}&
\mp(T_{\mu}N^{\sigma}-\Phi_{\mu\rho}S^{\rho}S^{\sigma})B^{\scriptscriptstyle{0}}_{\sigma\nu}
\pm B^{\scriptscriptstyle{0}}_{\mu\sigma}(N^{\sigma}T_{\nu}-S^{\sigma}S^{\rho}\Phi_{\rho\nu})\,.
\ea
\ee
After all, the DFT-metric becomes completely regular,
\be
\cH_{AB}=\left(\ba{cc}1&0\\B^{\scriptscriptstyle{0}}&1\ea\right)
\left(\ba{cc}\Upsilon-SS^{T}~&~
\pm\left(SS^{T}\Phi-NT^{T}\right)\\
\pm\left(\Phi S S^{T}-TN^{T}\right)~&~
\Phi-\Phi S S^{T}\Phi
\ea\right)\left(\ba{cc}1&-B^{\scriptscriptstyle{0}}\\0&1\ea\right)\,,
\ee
which can easily and precisely be identified as   the  $(1,1)$ type of the classification~(\ref{cHFINAL}) as
\be
\ba{lll}
\Upsilon^{\mu\nu}-S^{\mu}S^{\nu}\equiv H^{\mu\nu}\,,\quad&\quad \Phi_{\mu\nu}-\Phi_{\mu\rho}S^{\rho}\Phi_{\nu\sigma}
S^{\sigma}\equiv K_{\mu\nu}\,,\quad&\quad
\big\{T_{\mu}\,,\, \Phi_{\nu\rho}S^{\rho}\big\}\equiv\big\{X_{\mu}\,,\,\brX_{\nu}\big\}\,.
\ea
\ee
As  demonstrated in \cite{Lee:2013hma},  constant flat  background belonging to this type  generates   the Gomis-Ooguri non-relativistic string~\cite{Gomis:2000bd} (see also \cite{Park:2016sbw} for its Green-Schwarz superstring extension). Thus, a generic $(1,1)$ DFT-metric provides a curved spacetime  generalization of the   non-relativistic string.


\subsection{$(D-1,0)$\,: Ultra-relativistic or Carroll\label{SECDm1}}

The Riemannian metric~(\ref{gc}) in the previous section  defines  the \textit{proper length}.  Rescaling the metric by an overall factor of $c^{-2}$,  it becomes  the Riemannian metric for  the \textit{proper time}:
\be
\ba{l}
g_{\mu\nu}=-T_{\mu}T_{\nu}(1-S^{\rho}S^{\sigma} \Phi_{\rho\sigma})+\frac{2}{c}T_{(\mu}\Phi_{\nu)\rho}S^{\rho}+\frac{1}{c^{2}}\Phi_{\mu\nu}\,,\\
g^{\mu\nu}
=c^{2}\left(\Upsilon^{\mu\nu}-S^{\mu}S^{\nu}\right)+2cN^{(\mu}S^{\nu)}-N^{\mu}N^{\nu}\,,
\ea
\label{gc2}
\ee
where the variables should satisfy \eqref{TS} and \eqref{TNUpsilon}, which we recall here
\be
\ba{lllll}
T_{\mu}S^{\mu}=0\,,\quad&\quad
T_{\mu}N^{\mu}=1\,,\quad&\quad
T_{\mu}\Upsilon^{\mu\nu}=0\,,\quad&\quad
N^{\mu}\Phi_{\mu\nu}=0\,,\quad&\quad
T_{\mu}N^{\nu}+\Phi_{\mu\rho}\Upsilon^{\rho\nu}=\delta_{\mu}{}^{\nu}\,.
\ea
\label{TNUpsilon2}
\ee

Clearly, the expression of $g^{\mu\nu}$ in (\ref{gc2}) indicates the possibility of  taking  a  small $c$   \textit{i.e.~}ultra-relativistic  limit, as the inverse remains  non-singular, yet degenerate having rank one,
\be
\dis{\lim_{c\rightarrow 0}}\, g^{\mu\nu}=-N^{\mu}N^{\nu}\,.
\label{limgN}
\ee
In this subsection, we  propose   a $(D-1,0)$  DFT-metric  as  the ultra-relativistic `completion' of the above degenerate inverse~(\ref{limgN}),
\be
\cH_{AB}=\left(\ba{cc}1&0\\B&1\ea\right)
\left(\ba{cc}-NN^{T}~&~
\Upsilon\Phi\\
\Phi\Upsilon~&~
-TT^{T}
\ea\right)\left(\ba{cc}1&-B\\0&1\ea\right)\,,
\label{Dm1}
\ee
where all the variables are from \eqref{TNUpsilon2}. It is easy to check that this ansatz satisfies the defining properties of the DFT-metric~(\ref{cHdef}) and $\cH_{A}{}^{A}=2\Upsilon^{\mu\nu}\Phi_{\mu\nu}=2(D-1)$. Note the identification,
\be
\ba{lll}
H^{\mu\nu}\equiv-N^{\mu}N^{\nu}\,,\quad&\quad
K_{\mu\nu}\equiv-T_{\mu}T_{\nu}\,,\quad&\quad
\dis{\sum_{i=1}^{D-1}}\,X_{\mu}^{i}Y^{\nu}_{i}\equiv \Phi_{\mu\rho}\Upsilon^{\rho\nu}\,.
\ea
\ee

From (\ref{FROZEN}), particles freeze over almost all the directions except one,
\be
\Phi_{\mu\nu}\dot{x}^{\nu}\equiv 0\,. 
\label{almost}
\ee
This is in agreement with the ultra-relativistic limit of Riemannian geodesics \textit{\`a la}~Bergshoeff~\textit{et al.}~\cite{Bergshoeff:2014jla}. Namely,  particles  cannot move faster than light, and thus must freeze in the ultra-relativistic limit, $c\rightarrow 0$.

\noindent In fact,  $(D-1,0)$ forms a Carroll structure~\cite{Duval:2014uoa,Bekaert:2015xua}: $\Phi_{\mu\nu}$ is known as a \textit{Carrollian metric, i.e.}~a rank $(D-1)$ covariant metric whose kernel is spanned by the \textit{Carroll vector}, $N^{\nu}$, and  $T_{\mu}$ is a \textit{principal connection}. The \textit{Carrollian boost} symmetry ~\cite{Bekaert:2015xua} is given, with an arbitrary local parameter, $V^{\mu}$, by
\be
\ba{lll}
T_{\mu}\quad&\longmapsto&\quad T_{\mu}+\Phi_{\mu\nu}V^{\nu}\,,\\
\Upsilon^{\mu\nu}\quad&\longmapsto&\quad \Upsilon^{\mu\nu}-2N^{(\mu}\Upsilon^{\nu)\rho}\Phi_{\rho\sigma}V^{\sigma}+N^{\mu}N^{\nu}\Phi_{\rho\sigma}V^{\rho}V^{\sigma}\,,\\
B_{\mu\nu}\quad&\longmapsto&\quad B_{\mu\nu}+2T_{[\mu}\Phi_{\nu]\rho}V^{\rho}\,,
\ea
\ee
which leaves  our $(D-1,0)$ DFT-metric~(\ref{Dm1}) invariant, and can be identified with  the symmetry of the DFT-vielbein~(\ref{MilneG3}) for the case of $(D-1,0)$.


\subsection{Least non-Riemannian $(1,0)$ or $(0,1)$\,:  Non-relativistic or Newton-Cartan \label{SEC10}}
The ordinary Kaluza-Klein ansatz~(\ref{ordinaryKK}) treats the two block-diagonal Riemannian metrics, $g$ and $g^{\prime}$,  asymmetrically. Exchanging  the two   will lead to an alternative Kaluza-Klein ansatz.  In this subsection, we consider such an alternative ansatz for  the Riemannian metric~(\ref{gc}), which reads
\be
\ba{lll}
g_{\mu\nu}&=&(\delta_{\mu}{}^{\rho}-c^{-1}\Phi_{\mu\kappa}U^{\kappa}N^{\rho})
(\delta_{\nu}{}^{\sigma}-c^{-1}\Phi_{\nu\lambda}U^{\lambda}N^{\sigma})
(-c^{2}T_{\rho}T_{\sigma}+\Phi_{\rho\sigma})\\
{}&=&
-c^{2}T_{\mu}T_{\nu}+2cT_{(\mu}\Phi_{\nu)\rho}U^{\rho}
+\Phi_{\mu\nu}-\Phi_{\mu\rho}U^{\rho}\Phi_{\nu\sigma}U^{\sigma}\,,
\ea
\ee
with the inverse,
\be
\ba{lll}
g^{\mu\nu}&
=&
(\delta^{\mu}{}_{\rho}+c^{-1}N^{\mu}U^{\kappa}\Phi_{\kappa\rho})
(\delta^{\nu}{}_{\sigma}+c^{-1}N^{\nu}U^{\lambda}\Phi_{\lambda\sigma})
(-c^{-2}N^{\rho}N^{\sigma}+\Upsilon^{\rho\sigma})\\
{}&=&\Upsilon^{\mu\nu}+2c^{-1}N^{(\mu}U^{\nu)}-c^{-2}N^{\mu}N^{\nu}\left(1-U^{\rho}\Phi_{\rho\sigma}U^{\sigma}+2cT_{\rho}U^{\rho}\right)\,.
\ea
\ee
Clearly the inverse of the Riemannian metric allows non-singular large $c$ limit,
\be
\dis{\lim_{c\rightarrow\infty}}\,g^{\mu\nu}=\Upsilon^{\mu\nu}\,,
\ee
of which the rank is $D-1$.

The DFT-metric which completes this degenerate inverse is then 
\be
\cH_{AB}=\left(\ba{cc}1&0\\B&1\ea\right)
\left(\ba{cc}\Upsilon~&~\pm NT^{T}\\
\pm TN^{T}~&~\Phi
\ea\right)\left(\ba{cc}1&-B\\0&1\ea\right)\,,
\label{least}
\ee
with $\cH_{A}{}^{A}=\pm 2$. Here the upper and lower signs correspond to  $(1,0)$ and $(0,1)$ respectively.

Satisfying (\ref{TNUpsilon}) which we recall
\be
\ba{llll}
T_{\mu}N^{\mu}=1\,,\quad&\quad
T_{\mu}\Upsilon^{\mu\nu}=0\,,\quad&\quad
N^{\mu}\Phi_{\mu\nu}=0\,,\quad&\quad
T_{\mu}N^{\nu}+\Phi_{\mu\rho}\Upsilon^{\rho\nu}=\delta_{\mu}{}^{\nu}\,,
\ea
\ee
  $\{T_{\lambda}, \Upsilon^{\mu\nu}\}$ forms a Leibnizian structure (\cf \eg \cite{Bernal:2002ph,Bekaert:2014bwa}): $T_{\lambda}$ is the \textit{absolute clock} and $\Upsilon^{\mu\nu}$ is a collection of \textit{absolute rulers} with non-negative signature, \textit{i.e.~}$\eta_{ab}=\delta_{ab}$ from \eqref{HKhk}.  Further, the vector, $N^{\mu}$,  corresponds to a field of \textit{observers}, and the covariant rank ${D-1}$ metric, $\Phi_{\mu\nu}$, provides the associated  \textit{transverse metric}.  The transformation~\eqref{MilneG} reduces to
  \be
\ba{cll}
N^{\mu}~~&\longmapsto&~~ N^{\mu}+U^{\mu}\,,\\
\Phi_{\mu\nu}~~&\longmapsto&~~ \Phi_{\mu\nu}-2T_{(\mu}\Phi_{\nu)\rho}U^{\rho}
+U^{\rho}\Phi_{\rho\sigma}U^{\sigma}T_{\mu}T_{\nu}\,,\\
B_{\mu\nu}~~&\longmapsto&~~ B_{\mu\nu}\mp 2T_{[\mu}\Phi_{\nu]\rho}U^{\rho}\,,
\ea
\label{Milne}
\ee
with $U^{\mu}=\Upsilon^{\mu\nu}V_{\nu}\in \mbox{Ker}(T)$. This transformation is sometimes referred to as a Milne transformation or a Galilean boost in the literature~\cite{Duval:1993pe}.

\noindent From (\ref{FROZEN}), particles freeze over the time direction only, 
\be
T_{\mu}\dot{x}^{\mu}=0\,.
\label{frozenT}
\ee
so that the observer $\dot x^\mu$ is said to be \textit{space-like}. This is naturally  dual to the ultra-relativistic Carroll dynamics~\eqref{almost} where time flows but all spatial directions freeze.  
~\\

\noindent In order to account for the dynamics of \textit{time-like} observers (for which time flows), one needs to  introduce external forces, as done in the following subsection within the ambient framework  of a null Kaluza-Klein reduction.


\subsection{Embedding $(0,1)$ into ambient $(0,0)$ Kaluza-Klein ansatz\,:   Carroll or Newton-Cartan}

We start by considering  the $\hat{D}={1+D}$ Kaluza-Klein ansatz~(\ref{dramatic}) for a Riemannian  ambient DFT-metric, \textit{i.e.~}$(\hat{n},\hat{\brn})=(0,0)$.  As for the `internal' space, we assume   $D^{\prime}=1$, $(n^{\prime},\brn^{\prime})=(1,0)$  with  $\cH^{\prime}_{\Apr\Bpr}\equiv\cJpr_{\Apr\Bpr}$.   Then the `external' DFT-metric, $\cH_{AB}$, must be of the  $(n,\brn)=(0,1)$ type\footnote{The alternative choice of  $(n^{\prime},\brn^{\prime})=(0,1)$  obtained by setting $\cH^{\prime}_{\Apr\Bpr}\equiv-\cJpr_{\Apr\Bpr}$ will involve replacing  $\brP$ by $-P$ in (\ref{dramatic}), and accordingly  the external DFT-metric, $\cH$, will need to be of $(1,0)$ type.},  \textit{i.e.~}the lower sign in \eqref{least}, which ensures $\hcH_{\hA}{}^{\hA}=2(\hat{n}-\hat{\brn})=2(n+n^{\prime}-\brn-\brn^{\prime})=0$.   We let $(\ty,y)$ denote  the primed coordinates, $(\tx_{1}^{\prime},x^{\prime 1})$,  and write  for  the ambient doubled coordinates, 
\be
\rD_{\tau}x^{\hat{A}}=(\rD_{\tau}\ty\,,\,\dot{y}\,,\,\rD_{\tau}x^{A})=(\dot{\ty}-A_{\tau\ty}\,,\,\dot{y}\,,\,\dot{\tx}_{\mu}-A_{\tau\mu}\,,\,\dot{x}^{\nu})\,.
\ee   
We solve the constraint on $W_{M^{\prime}}{}^{N}$~(\ref{Wconstraint}) by putting  $W_{\mu^{\prime}}{}^{N}\equiv0$, such that for the present case of $D^{\prime}=1$,  we simply have
\be
W_{\Mpr}{}^{N}\equiv\left(\,W^{N},\,0\,\right)\,,
\label{Wansatz}
\ee
where the $\ODD$ vector, $W^{N}$, carries  no hidden index.
By choosing this -- instead of letting \textit{e.g.~}$W^{\mu^{\prime}N}$ vanish -- we ensure a \textit{null  Killing vector}, ${\xi}^{\hat{A}}=(\tilde{{\xi}}_{\hat{\mu}}\,,\,{\xi}^{\hat{\nu}})$~(\ref{deltaXYHK}), (\ref{gLie}) with $\xi^{\hat{\mu}}\partial_{\hat{\mu}}=\partial_{y}$, satisfying from (\ref{Length}),\footnote{In terms of ordinary Lie derivative, $\cL_{\xi}\hat{g}_{\hmu\hnu}=0$,   $\,\cL_{\xi}\hat{B}_{\hmu\hnu}=-2\partial_{[\hmu}\tilde{\xi}_{\hnu]}$,  \,and   (\ref{null}) means  $\xi^{\hmu}\xi^{\hnu}\hat{g}_{\hmu\hnu}=0$.}
\be
\ln\left[\int\cD\cA\,\exp\left(-\sqrt{(\xi^{\hA}-\cA^{\hA})(\xi^{\hB}-\cA^{\hB})\hcH_{\hA\hB}}\right)\right]=0\,.
\label{null}
\ee
The ambient DFT-metric~(\ref{dramatic}) then takes the following form, 
\be
\hcH_{\hA\hB}=\left(\ba{ccc}
-2W_{\brp}W^{\brp}\,~&~\,1\,~&~\,2\brV_{B\brp}W^{\brp}\\
1\,~&~\,0\,~&~\,0\\
2\brV_{A\brp}W^{\brp}\,~&~\,0\,~&~\,\cH_{AB}\ea\right)\,,
\label{dramatic2}
\ee
where, using the notations of section~\ref{SECvielbein},  we set a $(D+1)$-dimensional $\Spin(s+1,t+1)$ vector,\footnote{If we had chosen $(n,\brn)=(1,0)$, from (\ref{VbrV}),  the expression~\eqref{Wbrp}  would have reduced to `$W^{\brp}=W^{\bra}$'  without  $W_{\pm}$. }
\be
W^{\brp}= W^{A}\brV_{A}{}^{\brp}\equiv\left(\ba{ccc}
W^{\bra}\,,&\textstyle{\frac{1}{\sqrt{2}}}(W_{+}+W_{-})~,&\textstyle{\frac{1}{\sqrt{2}}}(W_{+}-W_{-})\ea\right)\,,
\label{Wbrp}
\ee 
such that, from (\ref{VbrV}),
\be
\ba{l}
\brP^{A}{}_{B}W^{B}=\brV^{A}{}_{\brp}W^{\brp}=\left(\ba{cc}
\textstyle{\frac{1}{\sqrt{2}}}\brk_{\mu\bra}W^{\bra}+T_{\mu}W_{-}+B_{\mu\rho}
\left(\textstyle{\frac{1}{\sqrt{2}}}W^{\brb}\brh_{\brb}{}^{\rho}+W_{+}N^{\rho}\right)~, & \textstyle{\frac{1}{\sqrt{2}}}W^{\brb}\brh_{\brb}{}^{\nu}
+W_{+}N^{\nu}\ea\right)\,,\\
W_{\brp}W^{\brp}=W_{A}W_{B}\brP^{AB}
=W_{\bra}W^{\bra}+2W_{+}W_{-}\,.
\ea
\ee

\noindent It is also convenient to define
from (\ref{cXcY}), (\ref{etas}), 
\be
W_{\mu}:=\sqrt{2}\brcX_{\mu}{}^{\brp}W_{\brp}=\sqrt{2}\brk_{\mu\bra}W^{\bra}+2W_{-}T_{\mu}\,.
\ee

\noindent Note the identification of the conventions,
\be
\ba{ll}
\Phi_{\mu\nu}\equiv K_{\mu\nu}=-\brk_{\mu}{}^{\bra}\brk_{\nu}{}^{\brb}\breta_{\bra\brb}\,,\quad&\quad
\Upsilon^{\mu\nu}\equiv H^{\mu\nu}=-\breta^{\bra\brb}\brh_{\bra}{}^{\mu}\brh_{\brb}{}^{\nu}\,.
\ea
\ee

\noindent Now,  with  the lower sign choice of (\ref{least}),  plugging \eqref{dramatic2} into the   master doubled-yet-gauged   action for a point particle~(\ref{particleaction}), we obtain  in a similar fashion to (\ref{DHD0}),

\be
\ba{lll}
S&=&\dis{\int\rd\tau~e^{-1\,}\rD_{\tau}x^{\hat{A}}\rD_{\tau}x^{\hat{B}}
\hat{\cH}_{\hat{A}\hat{B}}-\quarter m^{2}e}\\
{}&=&\dis{
\int\rd\tau~e^{-1}\Big[2\rD_{\tau}\ty\left(\dot{y}+2\rD_{\tau}x^{A}\brV_{A\brp}W^{\brp}-\rD_{\tau}\ty W_{\brp}W^{\brp}\right)
+\rD_{\tau}x^{A}\rD_{\tau}x^{B}\cH_{AB}\Big]-\quarter m^{2}e}\\
{}&=&\dis{\int\rd\tau}
\left[
\ba{l}
e^{-1}\Big[\dot{x}^{\mu}\dot{x}^{\nu}
\Phi_{\mu\nu}+2\rD_{\tau}\ty W_{\mu} \dot{x}^{\mu}-4(\rD_{\tau}\ty)^{2}W_{+}W_{-}
+2\dot{y}\rD_{\tau}\ty\Big]-\quarter m^{2}e\\
-2e^{-1}\left(T_{\mu}\dot{x}^{\mu}-2\rD_{\tau}\ty W_{+}\right)\Lambda
-e^{-1}\brh_{\bra}{}^{\mu}\Lambda_{\mu}\brh^{\bra\nu}\Lambda_{\nu}
\ea
\right]\,,
\ea
\label{twoexp}
\ee
where we set for shorthand notation as well as for a convenient field redefinition to replace  $A_{\tau\mu}$,
\be
\ba{ll}
\Lambda_{\mu}:=\dot{\tx}_{\mu}-A_{\tau\mu}-B_{\mu\kappa}\dot{x}^{\kappa}-\rD_{\tau}\ty W_{\mu}\,,\quad&\quad
\Lambda:=\Lambda_{\mu}N^{\mu}+2\rD_{\tau}\ty W_{-}\,.
\ea
\ee
Note that the very last term  in (\ref{twoexp}) is a perfect square which vanishes after  $\brh_{\bra}{}^{\mu}\Lambda_{\mu}$ being integrated out as
\be
\brh_{\bra}{}^{\mu}\Lambda_{\mu}\equiv0\,.
\label{hL}
\ee
Since $y$ is the coordinate for the  isometry direction, it serves as a Lagrange multiplier: it forces the  conjugate momentum of $y$, or $p$, to be constant,
\be
\ba{lll}
\frac{\rd~}{\rd\tau}\left(e^{-1}\rD_{\tau}\ty\right)\equiv0\quad&\quad\Longrightarrow\quad&\quad
2\rD_{\tau}\ty=ep\quad\mbox{with~~~\,\,constant~~~\,\,}p\,.
\ea
\label{c1}
\ee
Integrating out $\Lambda$ gives  a constraint,
\be
\cE_{\Lambda}:=T_{\mu}\dot{x}^{\mu}- ep W_{+}\equiv 0\,,
\label{c3}
\ee
such that the time is generically not frozen, \textit{c.f.~}\eqref{frozenT}. 
Further, integrating out the  auxiliary field, $A_{\tau\ty}$ inside $\rD_{\tau}\ty$ determines  the velocity,  with \eqref{hL}, \eqref{c1}, \eqref{c3},
\be
\ba{lll}
\dot{y}&=&epW_{\brp}W^{\brp}-2\rD_{\tau}x^{A}\brV_{A\brp}W^{\brp}\\
{}&=&-2 W_{+}\Lambda-W_{\mu}\dot{x}^{\mu}+2epW_{+}W_{-}\,.
\ea
\label{c2}
\ee

\noindent The einbein imposes the Hamiltonian constraint,
\be
\cE_{e}:=
\Phi_{\mu\nu}\dot{x}^{\mu}\dot{x}^{\nu}+e^{2}p^{2}W_{+}W_{-}- 2epW_{+}\Lambda+\quarter e^{2}m^{2}\equiv 0\,.
\label{einc}
\ee

\noindent From (\ref{c2}) and (\ref{einc}), it follows that: 
\be
-p\dot{y}=  e^{-1}\Phi_{\mu\nu}\dot{x}^{\mu}\dot{x}^{\nu}
+pW_{\mu}\dot{x}^{\mu}-ep^{2}W_{+}W_{-}+\quarter m^{2}e\,.
\ee
That is to say, whenever $p\neq0$, $\dot{y}$ is completely fixed by the  dynamics of the  $x^{\mu}$ coordinates. The auxiliary variable, $\Lambda$, is  also fixed in the same manner.\\

\noindent Making use of the world-line diffeomorphisms, we hereafter normalize the einbein:
\be
e\equiv 1\,,
\ee
such that $\tau$ coincides with the proper length.

\noindent The equation of motion for $x^{\mu}$ reads now
\be
\cE_{\mu}:=
\Phi_{\mu\nu}\ddot{x}^{\nu}+\left(
\partial_{\rho}\Phi_{\sigma\mu}-\half
\partial_{\mu}\Phi_{\rho\sigma}\right)
\dot{x}^{\rho}\dot{x}^{\sigma}
+\left(T_{\mu\nu}\Lambda-\half pW_{\mu\nu}\right)\dot{x}^{\nu}+\half p^{2}\partial_{\mu}(W_{+}W_{-})-p\Lambda\partial_{\mu}W_{+}-T_{\mu}\dot{\Lambda}\,,
\label{cEmu}
\ee
where we defined for simplicity, the field strengths
\be
\ba{ll}
T_{\mu\nu}:=\partial_{\mu}T_{\nu}-\partial_{\nu}T_{\mu}\,,\quad&\quad
W_{\mu\nu}:=\partial_{\mu}W_{\nu}-\partial_{\nu}W_{\mu}\,.
\ea
\ee
Computing the contractions, $N^{\mu}\cE_{\mu}$, $\dot{x}^{\mu}\cE_{\mu}$, respectively, we obtain the time derivative of the auxiliary variable,
\be
\dot{\Lambda}=N^{\mu}\Big[\left(
\partial_{\rho}\Phi_{\sigma\mu}-\half
\partial_{\mu}\Phi_{\rho\sigma}\right)
\dot{x}^{\rho}\dot{x}^{\sigma}
+\left(T_{\mu\nu}\Lambda-\half pW_{\mu\nu}\right)\dot{x}^{\nu}+\half p^{2}\partial_{\mu}(W_{+}W_{-})-p\Lambda\partial_{\mu}W_{+}\Big]\,,\label{eqdLambda}
\ee
and a consistency relation among   the constraints~(\ref{c3}), (\ref{einc}),
\be
\dot{x}^{\mu}\cE_{\mu}+\dot{\Lambda}\cE_{\Lambda}-
\half\dot{\cE}_{e}=0\,.
\ee
While \eqref{cEmu} determines partially the acceleration, $\ddot{x}^{\mu}$, the time derivative of the constraint~(\ref{c3}) can provide the missing component,
\be
\dot{\cE}_{\Lambda}=T_{\mu}\ddot{x}^{\mu}
+\partial_{(\mu}T_{\nu)}\dot{x}^{\mu}\dot{x}^{\nu}-p\dot{x}^{\mu}\partial_{\mu}W_{+}=0\,.
\label{dtcEL}
\ee
All together, the combination, $\Upsilon^{\lambda\mu}\cE_{\mu}+N^{\lambda}\dot{\cE}_{\Lambda}$, fully determines  the acceleration,
\be
\ddot{x}^{\lambda}+\gamma^{\lambda}_{\mu\nu}
\dot{x}^{\mu}\dot{x}^{\nu}+\left[\Upsilon^{\lambda\mu}
T_{\mu\nu}\Lambda-p(N^{\lambda}\partial_{\nu}W_{+}+
\half  \Upsilon^{\lambda\mu}W_{\mu\nu})\right]\dot{x}^{\nu}
+\half p^{2}\Upsilon^{\lambda\mu}\partial_{\mu}(W_{+}W_{-})-p\Lambda\Upsilon^{\lambda\mu}\partial_{\mu}W_{+}=0\,,
\label{accel}
\ee 
where  $\gamma^{\lambda}_{\mu\nu}$ denotes the following coefficients, 
 \be
\gamma^{\lambda}_{\mu\nu}:=N^{\lambda}\partial_{(\mu}T_{\nu)}+
\half\Upsilon^{\lambda\rho}\left(
\partial_{\mu}\Phi_{\nu\rho}+\partial_{\nu}\Phi_{\rho\mu}-
\partial_{\rho}\Phi_{\mu\nu}\right)\label{coeffgamma}
\,.
\ee
\noindent We emphasize that the dynamics of the $D$-dimensional coordinates $x^\mu$ as prescribed by \eqref{accel} is independent of the Kaluza-Klein direction, $y$. Geometrically, this means that one can interpret $x^\mu$ as coordinates on the \textit{quotient manifold} of the ambient spacetime by the light-like direction along the vector field,  $\xi^{\hat{\mu}}\partial_{\hat{\mu}}=\partial_{y}$. \\
~\\\noindent In the special case where  $T_{\mu\nu}$ vanishes (\ie~the one-form, $T_{\mu}$, is closed) and $W_{+}$ is a (non-vanishing) constant, the expression~(\ref{accel}) simplifies to
\be
\ddot{x}^{\lambda}+\gamma^{\lambda}_{\mu\nu}
\dot{x}^{\mu}\dot{x}^{\nu}=
\half p \Upsilon^{\lambda\mu}\left[W_{\mu\nu}\dot{x}^{\nu}
-p  \partial_{\mu}(W_{+}W_{-})\right]\,,\label{eqmotion}
\ee
of which the right-hand side can be interpreted as the Lorentz plus Coulomb forces. In this particular case, the  coefficients~\eqref{coeffgamma} are the ones associated to the so-called `special Galilean connection' for the field of observers, $N^{\mu}$, (\cf~\eg~\cite{Bekaert:2014bwa}). In accordance with the usual Riemannian ambient approach of \cite{Duval:1984cj, Duval:1990hj, Julia:1994bs} (\cf also \cite{Minguzzi:2006wz,Bekaert:2013fta,Bekaert:2015xua,BekaertSimons}), the resulting dynamical trajectories~\eqref{eqmotion} can be interpreted as Newton-Cartan geodesics. These are of two different types, depending on  the value of $p$\,:
\begin{itemize}
\item $p=0$\,  (Space-like observer).\\
In this case, the constraint\footnote{From the ambient perspective, the constraint, $T_\mu\dot x^{\mu}= 0$, implies that the dynamics becomes restricted to a $D$-dimensional hypersurface of the ambient manifold, transverse to the null isometry vector field, $\xi^{\hat{\mu}}\partial_{\hat{\mu}}=\partial_{y}$. Such a light-like hypersurface is naturally endowed with a Carrollian metric structure \cite{Duval:2014uoa}, and the equations of motion~\eqref{eqmotion} together with \eqref{c2} and \eqref{eqdLambda} can be naturally interpreted as  geodesics associated to a suitable {Carrollian} connection induced by the ambient metric structure (\cf \cite{Duval:2014uoa,Morand2017} for details). The role of Carrollian time is then played by ${y}$ and  the `space-like' directions are generically unfrozen, thus generalizing the Carrollian dynamics discussed in section~\ref{SECDm1}. }, $T_\mu\dot x^{\mu}= 0$, holds as a consequence of \eqref{c3} so that we recover the case investigated in section~\ref{SEC10} for which time freezes. Geometrically, the observer trajectory is restricted to a $(D-1)$-dimensional hypersurface (absolute space). The absolute spaces are Riemannian spaces (of Euclidean signature), since the degenerate metric $\Phi_{\mu\nu}$ becomes invertible on $\Ker T$ (\cf \eg \cite{Bekaert:2014bwa}). Equation~\eqref{eqmotion} thus describes geodesics associated to the spatial Riemannian metric and the Hamiltonian  constraint~\eqref{einc} can be solved as $e=\frac{2}{|m|}\sqrt{\Phi_{\mu\nu}\dot{x}^{\mu}\dot{x}^{\nu}}\equiv1$. 
\\

\item $p\neq0$ (Time-like observer).

\noindent In this case, $\tau$ is parametrized to ensure $e=\frac{1}{pW_+}T_\mu\dot x^\mu\equiv1$ such that the observer $\dot x^\mu$ is time-like. Equation \eqref{eqmotion} can thus be reformulated as
\be
\ddot{x}^{\lambda}+\hat{\gamma}^{\lambda}_{\mu\nu}
\dot{x}^{\mu}\dot{x}^{\nu}=0\, ,
\ee
where the coefficients $\hat{\gamma}^{\lambda}_{\mu\nu}$ are defined as
\bea
\hat{\gamma}^{\lambda}_{\mu\nu}:=\gamma^{\lambda}_{\mu\nu} +\Upsilon^{\lambda\rho}\, T_{(\mu}F_{\nu)\rho}\, ,\label{coeffNewt}
\eea
with $F_{\mu\nu}:=\p_\mu A_\nu-\p_\nu A_\mu$ and $A_\mu:=\frac{1}{2\, W_+}\pl W_\mu-W_-\, T_\mu\pr$. ~\\
The connection associated to the coefficients \eqref{coeffNewt} is naturally interpreted as a Newtonian connection \cite{Kuenzle:1972zw}, \ie a torsion-free connection compatible with the Leibnizian structure $(\Upsilon^{\mu\nu},T_\mu)$ such that the associated field strength, $F_{\mu\nu}$, is closed. 
\end{itemize}
\noindent In summary, assuming the triviality of $T_{\mu\nu}$ and $W_{+}$, the doubled-yet-gauged  particle action~\eqref{particleaction} with the ambient  $(D+1)$-dimensional Kaluza-Klein ansatz~\eqref{dramatic} reproduces the full content of  Newtonian dynamics (unifying the space-like and time-like cases) on the $D$-dimensional  manifold quotient along the light-like direction, $y$. 

\noindent In principle,  the assumption regarding the triviality of the variables, $T_{\mu\nu}$ and $W_{+}$, should be examined by considering the on-shell dynamics of the DFT-metric, \textit{i.e.~}the Euler-Lagrangian equations of DFT. 
In the present work, we have focused on the kinematical side of the DFT-metric and the subsequent particle and string dynamics on the background. We leave the investigation of the dynamical aspect of the $(n,\brn)$ DFT-metric for future work.  From our perspective, the DFT action  and its full equations of motion determine universally and unambiguously all the dynamics of the $(n,\brn)$ backgrounds, including $(0,0)$ Riemannian General Relativity and $(0,1)$ Newton-Cartan gravity, in a unifying manner. 
\section*{Acknowledgements}
We  would like to thank Xavier Bekaert and Daniel Waldram for helpful discussions as well as  Soo-Jong Rey for supportive encouragement.    
K.M is grateful to Sogang University  for  hospitality while  part of this work was completed.  
This work was  supported by  the National Research Foundation of Korea   through  the Grants   2015K1A3A1A21000302  and  2016R1D1A1B01015196, as well as by the Chilean Fondecyt Postdoc Project $\text{N}^\circ$3160325.
\\


\newpage
\appendix
\section{Derivation of the most general form of the DFT-metric, \,Eq.(\ref{cHFINAL})}

By definition~(\ref{cHdef}), the DFT-metric is a symmetric $\ODD$ element, such that it   satisfies
\be
\ba{ll}
\cH_{MN}=\cH_{NM}\,,\quad&\quad \cH_{L}{}^{M}\cH_{M}{}^{N}=\delta_{L}{}^{N}\,.
\ea
\label{defcH}
\ee
With respect to the $\ODD$ metric~(\ref{cJ}) and the choice of the section, $\tpartial^{\mu}\equiv 0$, we decompose the DFT-metric, 
\be
\cH_{MN}=\left(\ba{cc}\cH^{\mu\nu}&\cH^{\mu}{}_{\lambda}\\
\cH_{\kappa}{}^{\nu}&\cH_{\kappa\lambda}\ea\right)\,.
\ee
The defining condition~(\ref{defcH}) reads  explicitly,
\be
\ba{lll}
\cH^{\mu\nu}=\cH^{\nu\mu}\,,\quad&\quad\cH_{\mu\nu}=\cH_{\nu\mu}\,,\quad&\quad
\cH_{\mu}{}^{\nu}=\cH^{\nu}{}_{\mu}\,,\\
\cH^{(\mu}{}_{\rho}\cH^{\nu)\rho}=0\,,\quad&\quad
\cH_{\rho(\mu}\cH^{\rho}{}_{\nu)}=0\,,\quad&\quad
\cH^{\mu}{}_{\rho}\cH^{\rho}{}_{\nu}+\cH^{\mu\rho}\cH_{\rho\nu}=\delta^{\mu}{}_{\nu}\,.
\ea
\label{defcH2}
\ee
The generalized Lie derivative of the DFT-metric, \textit{c.f.~}(\ref{diff2}),
\be
\hcL_{\xi}\cH_{AB}=\xi^{C}\partial_{C}\cH_{AB}+(\partial_{A}\xi_{C}-\partial_{C}\xi_{A})\cH^{C}{}_{B}+
(\partial_{B}\xi_{C}-\partial_{C}\xi_{B})\cH_{A}{}^{C}\,,
\label{gLie}
\ee
leads to
\be
\ba{ll}
\delta\cH^{\mu\nu}=\cL_{\xi}\cH^{\mu\nu}\,,\quad&\quad
\delta\cH_{\kappa\lambda}=\cL_{\xi}\cH_{\kappa\lambda}+(\partial_{\kappa}\tilde{\xi}_{\rho}-\partial_{\rho}\tilde{\xi}_{\kappa})\cH^{\rho}{}_{\lambda}-\cH_{\kappa}{}^{\rho}(\partial_{\rho}\tilde{\xi}_{\lambda}-\partial_{\lambda}\tilde{\xi}_{\rho})\,,\\
\delta\cH^{\mu}{}_{\lambda}=\cL_{\xi}\cH^{\mu}{}_{\lambda}-\cH^{\mu\rho}(\partial_{\rho}\tilde{\xi}_{\lambda}-\partial_{\lambda}\tilde{\xi}_{\rho})\,,\quad&\quad
\delta\cH_{\kappa}{}^{\nu}=\cL_{\xi}\cH_{\kappa}{}^{\nu}+(\partial_{\kappa}\tilde{\xi}_{\rho}-\partial_{\rho}\tilde{\xi}_{\kappa})\cH^{\rho\nu}\,.
\ea
\label{GLD2}
\ee  
  
  Viewed as a  $D\times D$ matrix, if $\cH^{\mu\nu}$ is non-degenerate, we may identify it as the inverse of a Riemannian metric. It is easy to see then that the remaining constraints are all  solved by a skew-symmetric $B$-field, such that the most general  DFT-metric  in this case takes the well-known form~(\ref{RiemanncH}). Henceforth, we  look for the most general form of the DFT-metric, where $\cH^{\mu\nu}$ is degenerate. Firstly, we  focus on the case where the rank of $\cH^{\mu\nu}$ is $D{-1}$, admitting  only one zero-eigenvector, $X_{\mu}$,
\be
\ba{ll}
\cH^{\mu\nu}\equiv H^{\mu\nu}\,,\quad&\quad H^{\mu\nu}X_{\nu}=0\,.
\ea
\ee
From (\ref{defcH2}), $\cH^{\mu}{}_{\rho}H^{\rho\nu}$ is skew-symmetric, and hence
\be 
X_{\mu}\cH^{\mu}{}_{\rho}H^{\rho\nu}=-\cH^{\nu}{}_{\rho}H^{\rho\mu}X_{\mu}=0\,.
\ee 
Without loss of generality then,  introducing  a skew-symmetric $B$-field,\footnote{The ambiguity in introducing the $B$-field through  \eqref{introB}  amounts to  the symmetry of the final result~\eqref{MilneG2}.} we may put
\be
\ba{ll}
\cH^{\mu}{}_{\rho}H^{\rho\nu}\equiv- H^{\mu\rho}B_{\rho\sigma}H^{\sigma\nu}\,,\quad&\quad
{B}_{\mu\nu}=-{B}_{\nu\mu}\,.
\ea
\label{introB}
\ee
It follows that, with some vector field,    $Y^{\mu}$, $\cH^{\mu}{}_{\nu}$ takes the form,
\be
\cH^{\mu}{}_{\nu}=-H^{\mu\rho}B_{\rho\nu}
+Y^{\mu}X_{\nu}\,.
\label{cH12}
\ee 
We proceed with a new symmetric variable, $K_{\mu\nu}=K_{\nu\mu}$,
\be
\cH_{\mu\nu}\equiv K_{\mu\nu}
-B_{\mu\rho}H^{\rho\sigma}B_{\sigma\nu}
+2X_{(\mu}B_{\nu)\rho}Y^{\rho}\,.
\label{cH22}
\ee
The last relation in (\ref{defcH2}) gives
\be
H^{\mu\rho}K_{\rho\nu}
+(Y^{\rho}X_{\rho})Y^{\mu}X_{\nu}=\delta^{\mu}{}_{\nu}\,.
\label{hg1}
\ee
Contracting this with $X_{\mu}$   shows 
\be
Y^{\mu}X_{\mu}=\pm 1\,.
\label{YX}
\ee  
Lastly we impose the  skew-symmetric condition of $\cH_{\mu\rho}\cH^{\rho}{}_{\nu}$, which gives with (\ref{hg1}), 
\be
K_{\mu\rho}Y^{\rho}X_{\nu}
+K_{\nu\rho}Y^{\rho}X_{\mu}=0\,,
\label{gNt}
\ee
and hence in particular, contracting with $Y^{\mu}$, we have
\be
K_{\nu\rho}Y^{\rho}=
\mp(Y^{\mu}K_{\mu\rho}Y^{\rho})X_{\nu}\,.
\label{gN}
\ee
Substituting this back into  (\ref{gNt}), we conclude that  $Y^{\mu}K_{\mu\rho}Y^{\rho}$ must  be trivial, and hence in fact from (\ref{gN}),
\be
K_{\nu\rho}Y^{\rho}=0\,.
\ee
We may perform a field redefinition, $Y^{\mu}\rightarrow \pm Y^{\mu}$, in order to remove the sign factor in the normalization of (\ref{YX}). After all, the most general form of the DFT-metric in the `least'   degenerate case  takes the form,
\be
\ba{ll}
\cH_{MN}=\left(\ba{cc}H^{\mu\nu}&
-H^{\mu\sigma}B_{\sigma\lambda}\pm Y^{\mu}X_{\lambda}\\
B_{\kappa\rho}H^{\rho\nu}\pm X_{\kappa}Y^{\nu}&~~
K_{\kappa\lambda}-B_{\kappa\rho}H^{\rho\sigma}B_{\sigma\lambda}
\pm 2X_{(\kappa}B_{\lambda)\rho}Y^{\rho}\ea\right)\,,
\ea
\label{LEAST}
\ee
of which the variables must meet
\be
\ba{lllll}
H^{\mu\nu}X_{\nu}=0\,,\quad&~~~ K_{\mu\nu}Y^{\nu}=0\,,\quad&~~~ Y^{\mu}X_{\mu}= 1\,,\quad&~~~ H^{\lambda\mu}K_{\mu\nu}+ Y^{\lambda}X_{\nu}=\delta^{\lambda}{}_{\nu}\,,\quad&~~~ B_{\mu\nu}=-B_{\nu\mu}\,.
\ea
\ee

\noindent The above analysis can be   straightforwardly   extended  to the most general  degenerate cases, where there are $N$ number of linearly independent zero-eigenvectors, $X^{i}_{\mu}$, $i=1,2,\cdots, N$, such that the rank of   $\cH^{\mu\nu}$ is $D-N$, 
\be
\ba{ll}
\cH^{\mu\nu}\equiv H^{\mu\nu}\,,\qquad&\qquad H^{\mu\nu}X^{i}_{\nu}=0\,.
\ea
\ee
From
\be
\ba{lll}
\cH^{(\mu}{}_{\rho}H^{\nu)\rho}=0\,,\quad&\quad \cH^{\mu}{}_{\rho}H^{\rho\nu}X^{i}_{\nu}=0\,,
\quad&\quad X^{i}_{\mu}\cH^{\mu}{}_{\rho}H^{\rho\nu}=0\,,
\ea
\ee 
Eqs.(\ref{cH12}) and (\ref{cH22}) generalize, defining   $Y_{i}^{\mu}$ and $M_{\mu\nu}$,  to
\be
\ba{ll}
\cH^{\mu}{}_{\nu}\equiv-H^{\mu\rho}B_{\rho\sigma}
+Y_{i}^{\mu}X^{i}_{\nu}\,,\quad&\quad
\cH_{\mu\nu}\equiv
M_{\mu\nu}-B_{\mu\rho}H^{\rho\sigma}B_{\sigma\nu}
+2X^{i}_{(\mu}B_{\nu)\rho}Y_{i}^{\rho}\,,
\ea
\ee 
such that the DFT-metric  assumes the following  intermediate form,
\be
\cH_{MN}=\left(\ba{cc}H^{\mu\nu}&
-H^{\mu\sigma}B_{\sigma\lambda}+Y_{i}^{\mu}X^{i}_{\lambda}\\
B_{\kappa\rho}H^{\rho\nu}+X^{i}_{\kappa}Y_{i}^{\nu}\quad&\quad
~~M_{\kappa\lambda}-B_{\kappa\rho}H^{\rho\sigma}B_{\sigma\lambda}
+2X^{i}_{(\kappa}B_{\lambda)\rho}Y_{i}^{\rho}
\ea\right)\,.
\label{intermediate}
\ee
In the above, the repeated index, $i$,  is summed from $1$ to $N$. The remaining  constraints in (\ref{defcH2}) give
\be
H^{\mu\rho}M_{\rho\nu}
+(Y_{i}^{\rho}X^{j}_{\rho})Y_{j}^{\mu}X^{i}_{\nu}=\delta^{\mu}{}_{\nu}\,,
\label{hg1G}
\ee
\be
M_{\mu\rho}Y_{i}^{\rho}X^{i}_{\nu}+M_{\nu\rho}Y_{i}^{\rho}X^{i}_{\mu}=0\,.~~
\label{gNtG}
\ee
Contraction  of (\ref{hg1G}) with  $X^{k}_{\mu}$    leads to
\be
X^{i}_{\nu}(
Y_{i}{\cdot X}^{j}\,Y_{j}{\cdot X}^{k})=X^{k}_{\nu}\,,
\ee
where we set  $Y_{i}{\cdot X}^{j}\equiv Y_{i}^{\mu}X^{j}_{\mu}$. Since $k=1,2,\cdots, N$ is arbitrary and the $X^{k}_{\nu}$ are independent, the above result actually  implies  that $Y_{i}{\cdot X}^{j}$ is an  involutory $N\times N$ matrix, 
\be
Y_{i}{\cdot X}^{j}\,Y_{j}{\cdot X}^{k}=\delta_{i}{}^{k}\,.
\label{Ntdelta}
\ee
On the other hand,  contraction of (\ref{gNtG}) with $(Y_{j}{\cdot X}^{k})Y_{k}^{\mu}$ leads to
\be
M_{\nu\rho}Y_{j}^{\rho}=-(Y_{j}{\cdot X}^{k})
(Y_{k}{\cdot M}{\cdot Y}_{i})X^{i}_{\nu}\,,
\label{gNG}
\ee
where we set $Y_{k}{\cdot M}{\cdot Y}_{i}\equiv Y_{k}^{\rho}M_{\rho\sigma}Y_{i}^{\sigma}$ for shorthand notation. 
Substituting  into  (\ref{gNtG}), we get
\be
(Y_{j}{\cdot X}^{k})
(Y_{k}{\cdot M}{\cdot Y}_{i})
X^{i}_{\mu}X^{j}_{\nu}+(Y_{j}{\cdot X}^{k})
(Y_{k}{\cdot M}{\cdot Y}_{i})
X^{i}_{\nu}X^{j}_{\mu}=0\,,
\ee
which,   after contraction with $Y^{\mu}_{l}Y^{\nu}_{m}$ and  from (\ref{Ntdelta}), can be seen to be equivalent to
\be
(Y_{l}{\cdot X}^{i})(Y_{i}{\cdot M}{\cdot Y}_{m})=-(Y_{m}{\cdot X}^{i})(Y_{i}{\cdot M}{\cdot Y}_{l})\,.
\label{NNN}
\ee
It follows from (\ref{hg1G}), (\ref{Ntdelta}), (\ref{gNG}) that   $(Y_{i}{\cdot X}^{j})Y_{j}^{\mu}X^{i}_{\nu}$ and $H^{\mu\rho}M_{\rho\nu}$ are mutually orthogonal  and complementary~(\ref{hg1G})  projection matrices, 
\be
\ba{ll}
(Y_{i}{\cdot X}^{j})Y_{j}^{\lambda}X^{i}_{\mu}\,(Y_{k}{\cdot X}^{l})Y_{l}^{\mu}X^{k}_{\nu}=(Y_{i}{\cdot X}^{j})Y_{j}^{\lambda}X^{i}_{\nu}\,,\quad&\quad
H^{\lambda\rho}M_{\rho\mu}\,H^{\mu\sigma}M_{\sigma\nu}=
H^{\lambda\rho}M_{\rho\nu}\,,\\
(Y_{i}{\cdot X}^{j})Y_{j}^{\lambda}X^{i}_{\mu}\,
H^{\mu\sigma}M_{\sigma\nu}=0\,,\quad&\quad
H^{\lambda\rho}M_{\rho\mu}\,(Y_{k}{\cdot X}^{l})Y_{l}^{\mu}X^{k}_{\nu}=0\,.
\ea
\ee
Now, we may recast (\ref{gNG}) into
\be
\left[M_{\mu\nu}+X^{i}_{\mu}\left\{X^{l}_{\nu}(Y_{l}{\cdot X}^{k})
(Y_{k}{\cdot M}{\cdot Y}_{j})
X^{j}_{\rho}\right\}Y^{\rho}_{i}\right]Y^{\nu}_{m}=0\,.
\label{MMY}
\ee
It is crucial to note, from  the symmetric property, $Y_{k}{\cdot M}{\cdot Y}_{j}=Y_{j}{\cdot M}{\cdot Y}_{k}$,  that  the  free  indices, $\mu$ and $\nu$, below are symmetric,
\be
X^{i}_{\mu}\left\{X^{l}_{\nu}(Y_{l}{\cdot X}^{k})
(Y_{k}{\cdot M}{\cdot Y}_{j})
X^{j}_{\rho}\right\}Y^{\rho}_{i}=
X^{i}_{\nu}\left\{X^{l}_{\mu}(Y_{l}{\cdot X}^{k})
(Y_{k}{\cdot M}{\cdot Y}_{j})
X^{j}_{\rho}\right\}Y^{\rho}_{i}\,,
\ee
and further from the skew-symmetric property~(\ref{NNN}), that the free  indices, $\nu$ and $\rho$,  below  are skew-symmetric,
\be
X^{l}_{\nu}(Y_{l}{\cdot X}^{k})
(Y_{k}{\cdot M}{\cdot Y}_{j})
X^{j}_{\rho}=-X^{l}_{\rho}(Y_{l}{\cdot X}^{k})
(Y_{k}{\cdot M}{\cdot Y}_{j})
X^{j}_{\nu}\,.
\ee
Therefore, if we perform a field redefinition,
\be
\ba{lll}
B_{\mu\nu}~&~\longrightarrow~&~
B_{\mu\nu}+\half X^{i}_{\mu}(Y_{i}{\cdot X}^{k})
(Y_{k}{\cdot M}{\cdot Y}_{j})X^{j}_{\nu}\,,
\ea
\ee
among the components of the DFT-metric spelled in (\ref{intermediate}), 
$H^{\mu\sigma}B_{\sigma\lambda}$, $B_{\kappa\rho}H^{\rho\sigma}$, and $B_{\kappa\rho}H^{\rho\sigma}B_{\sigma\lambda}$ remain invariant, but $M_{\kappa\lambda}
+2X^{i}_{(\kappa}B_{\lambda)\rho}Y_{i}^{\rho}$ transforms as follows,
\be
\ba{lll}
M_{\kappa\lambda}
+2X^{i}_{(\kappa}B_{\lambda)\rho}Y_{i}^{\rho}
~&~\longrightarrow~&~
M_{\kappa\lambda}+X^{i}_{(\kappa}\left\{
X_{\lambda)}^{j}(Y_{j}{\cdot X}^{k})(Y_{k}{\cdot M}{\cdot Y}_{l})X^{l}_{\rho}
\right\}Y_{i}^{\rho}
+2X^{i}_{(\kappa}B_{\lambda)\rho}Y_{i}^{\rho}\,.
\ea
\ee
We then let
\be
K_{\kappa\lambda}:=M_{\kappa\lambda}+X^{i}_{(\kappa}\left\{
X_{\lambda)}^{j}(Y_{j}{\cdot X}^{k})(Y_{k}{\cdot M}{\cdot Y}_{l})X^{l}_{\rho}
\right\}Y_{i}^{\rho}\,,
\ee
which  nicely satisfies
\be
\ba{ll}
K_{\kappa\lambda}Y^{\lambda}_{i}=0\,,
\quad&\quad
H^{\mu\rho}K_{\rho\nu}
+(Y_{i}{\cdot X}^{j})Y_{j}^{\mu}X^{i}_{\nu}=\delta^{\mu}{}_{\nu}\,.
\ea
\label{NICE}
\ee
Finally, we  perform a similarity transformation, $(X^{i}_{\mu},Y_{j}^{\nu})\rightarrow (S^{i}{}_{k}X^{k}_{\mu}, Y_{k}^{\nu}S^{-1k}{}_{j})$,  which leaves $Y^{\mu}_{i}X^{i}_{\nu}$ invariant but diagonalizes $Y^{\rho}_{i}X^{j}_{\rho}$ with the eigenvalues of either $+1$ or $-1$. We   then let  $N=n+\brn$ in order to denote the numbers of the $+1$ and $-1$ eigenvalues of $Y^{\rho}_{i}X^{j}_{\rho}$.  If the corresponding eigenvalue is $-1$, we further perform a  field redefinition, $(\brX^{\bri}_{\mu},\brY_{\bri}^{\nu}):=(X^{i}_{\mu}, -Y_{i}^{\nu})$, which  involves   the  change of  the index from $i$ to $\bri$.   In this way, we arrive at the most general form of the DFT-metric, \eqref{cHFINAL},  classified by two non-negative integers, $n,\brn$. \\

\noindent It is also worthwhile to decompose the $B$-field  utilizing  the completeness relation~(\ref{COMP}),
\be
B_{\mu\nu}=\beta_{\mu\nu}+B_{\mu j}X^{j}_{\nu}-B_{\nu j}X^{j}_{\mu}+\brB_{\mu \brj}\brX^{\brj}_{\nu}-\brB_{\nu \brj}\brX^{\brj}_{\mu}+X^{i}_{\mu}X^{j}_{\nu}b_{ij}
+\brX^{\bri}_{\mu}\brX^{\brj}_{\nu}b_{\bri\brj}
+2X^{i}_{[\mu}\brX^{\brj}_{\nu]}b_{i\brj}\,,
\label{BDECOMP}
\ee
for which we set 
\be
\ba{lll}
\beta_{\mu\nu}:=(KH)_{\mu}{}^{\rho}(KH)_{\nu}{}^{\sigma}B_{\rho\sigma}\,,\quad&\quad
b_{ij}:=Y_{i}^{\mu}Y_{j}^{\nu}B_{\mu\nu}\,,\quad&\quad 
B_{\mu i}:=B_{\mu\nu}Y^{\nu}_{i}-X_{\mu}^{j}b_{ji}+\brX_{\mu}^{\brj}b_{i\brj}\,,\\
b_{i\brj}:=Y_{i}^{\mu}\brY_{\brj}^{\nu}B_{\mu\nu}\,,\quad&\quad
b_{\bri\brj}:=\brY_{\bri}^{\mu}\brY_{\brj}^{\nu}B_{\mu\nu}\,,\quad&\quad
\brB_{\mu \bri}:=B_{\mu\nu}\brY^{\nu}_{\bri}-\brX_{\mu}^{\brj}b_{\brj\bri}-X_{\mu}^{j}b_{j\bri}\,.
\ea
\ee
The variables, $B_{\mu i}$, $\brB_{\mu\bri}$ and $\beta_{\mu\nu}$ are completely orthogonal to the vectors, $Y^{\mu}_{j}$ and $\brY^{\mu}_{\brj}$,
\be
\ba{llllll}
B_{\mu i}Y^{\mu}_{j}=0\,,\quad&~~
B_{\mu i}\brY^{\mu}_{\brj}=0\,,\quad&~~
\brB_{\mu \hati}Y^{\mu}_{j}=0\,,\quad&~~
\brB_{\mu \hati}\brY^{\mu}_{\brj}=0\,,\quad&~~
\beta_{\mu\nu}Y^{\mu}_{j}=0\,,\quad&~~
\beta_{\mu\nu}\brY^{\mu}_{\brj}=0\,.
\ea
\ee


\newpage



\begin{thebibliography}{99}


\bibitem{Buscher:1987sk}
  T.~H.~Buscher,
  ``A Symmetry of the String Background Field Equations,''
  Phys.\ Lett.\ B {\bf 194} (1987) 59.
  doi:10.1016/0370-2693(87)90769-6



\bibitem{Buscher:1987qj}
  T.~H.~Buscher,
  ``Path Integral Derivation of Quantum Duality in Nonlinear Sigma Models,''
  Phys.\ Lett.\ B {\bf 201} (1988) 466.
  doi:10.1016/0370-2693(88)90602-8



\bibitem{Duff:1989tf}
  M.~J.~Duff,
  ``Duality Rotations in String Theory,''
  Nucl.\ Phys.\ B {\bf 335} (1990) 610.
  doi:10.1016/0550-3213(90)90520-N



\bibitem{Tseytlin:1990nb}
  A.~A.~Tseytlin,
  ``Duality Symmetric Formulation of String World Sheet Dynamics,''
  Phys.\ Lett.\ B {\bf 242} (1990) 163.
  doi:10.1016/0370-2693(90)91454-J



\bibitem{Tseytlin:1990va}
  A.~A.~Tseytlin,
  ``Duality symmetric closed string theory and interacting chiral scalars,''
  Nucl.\ Phys.\ B {\bf 350} (1991) 395.
  doi:10.1016/0550-3213(91)90266-Z



\bibitem{Hull:2004in}
  C.~M.~Hull,
  ``A Geometry for non-geometric string backgrounds,''
  JHEP {\bf 0510} (2005) 065
  doi:10.1088/1126-6708/2005/10/065
  [hep-th/0406102].



\bibitem{Hull:2006qs}
  C.~M.~Hull,
  ``Global aspects of T-duality, gauged sigma models and T-folds,''
  JHEP {\bf 0710} (2007) 057
  doi:10.1088/1126-6708/2007/10/057
  [hep-th/0604178].



\bibitem{Hull:2006va}
  C.~M.~Hull,
  ``Doubled Geometry and T-Folds,''
  JHEP {\bf 0707} (2007) 080
  doi:10.1088/1126-6708/2007/07/080
  [hep-th/0605149].



\bibitem{Siegel:1993xq}
  W.~Siegel,
  ``Two vierbein formalism for string inspired axionic gravity,''
  Phys.\ Rev.\ D {\bf 47} (1993) 5453
  doi:10.1103/PhysRevD.47.5453
  [hep-th/9302036].



\bibitem{Siegel:1993th}
  W.~Siegel,
  ``Superspace duality in low-energy superstrings,''
  Phys.\ Rev.\ D {\bf 48} (1993) 2826
  doi:10.1103/PhysRevD.48.2826
  [hep-th/9305073].



\bibitem{Hull:2009mi}
  C.~Hull and B.~Zwiebach,
  ``Double Field Theory,''
  JHEP {\bf 0909} (2009) 099
  doi:10.1088/1126-6708/2009/09/099
  [arXiv:0904.4664 [hep-th]].



\bibitem{Hull:2009zb}
  C.~Hull and B.~Zwiebach,
  ``The Gauge algebra of double field theory and Courant brackets,''
  JHEP {\bf 0909} (2009) 090
  doi:10.1088/1126-6708/2009/09/090
  [arXiv:0908.1792 [hep-th]].



\bibitem{Hohm:2010pp}
  O.~Hohm, C.~Hull and B.~Zwiebach,
  ``Generalized metric formulation of double field theory,''
  JHEP {\bf 1008} (2010) 008
  doi:10.1007/JHEP08(2010)008
  [arXiv:1006.4823 [hep-th]].


\bibitem{Aldazabal:2013sca}
  G.~Aldazabal, D.~Marques and C.~Nunez,
  ``Double Field Theory: A Pedagogical Review,''
  Class.\ Quant.\ Grav.\  {\bf 30} (2013) 163001
  doi:10.1088/0264-9381/30/16/163001
  [arXiv:1305.1907 [hep-th]].


\bibitem{Berman:2013eva}
  D.~S.~Berman and D.~C.~Thompson,
  ``Duality Symmetric String and M-Theory,''
  Phys.\ Rept.\  {\bf 566} (2014) 1
  doi:10.1016/j.physrep.2014.11.007
  [arXiv:1306.2643 [hep-th]].
  
\bibitem{Hohm:2013bwa}
  O.~Hohm, D.~L\"ust and B.~Zwiebach,
  ``The Spacetime of Double Field Theory: Review, Remarks, and Outlook,''
  Fortsch.\ Phys.\  {\bf 61} (2013) 926
  doi:10.1002/prop.201300024
  [arXiv:1309.2977 [hep-th]].


\bibitem{Park:2013mpa}
  J.~H.~Park,
  ``Comments on double field theory and diffeomorphisms,''
  JHEP {\bf 1306} (2013) 098
  doi:10.1007/JHEP06(2013)098
  [arXiv:1304.5946 [hep-th]].



\bibitem{Jeon:2011cn}
  I.~Jeon, K.~Lee and J.~H.~Park,
  ``Stringy differential geometry, beyond Riemann,''
  Phys.\ Rev.\ D {\bf 84} (2011) 044022
  doi:10.1103/PhysRevD.84.044022
  [arXiv:1105.6294 [hep-th]].



\bibitem{Hohm:2011si}
  O.~Hohm and B.~Zwiebach,
  ``On the Riemann Tensor in Double Field Theory,''
  JHEP {\bf 1205} (2012) 126
  doi:10.1007/JHEP05(2012)126
  [arXiv:1112.5296 [hep-th]].



\bibitem{Lee:2013hma}
  K.~Lee and J.~H.~Park,
  ``Covariant action for a string in "doubled yet gauged" spacetime,''
  Nucl.\ Phys.\ B {\bf 880} (2014) 134
  doi:10.1016/j.nuclphysb.2014.01.003
  [arXiv:1307.8377 [hep-th]].



\bibitem{Ko:2015rha}
  S.~M.~Ko, C.~Melby-Thompson, R.~Meyer and J.~H.~Park,
  ``Dynamics of Perturbations in Double Field Theory \& Non-Relativistic String Theory,''
  JHEP {\bf 1512} (2015) 144
  doi:10.1007/JHEP12(2015)144
  [arXiv:1508.01121 [hep-th]].



\bibitem{Park:2016sbw}
  J.~H.~Park,
  ``Green-Schwarz superstring on doubled-yet-gauged spacetime,''
  JHEP {\bf 1611} (2016) 005
  doi:10.1007/JHEP11(2016)005
  [arXiv:1609.04265 [hep-th]].



\bibitem{Gomis:2000bd}
  J.~Gomis and H.~Ooguri,
  ``Nonrelativistic closed string theory,''
  J.\ Math.\ Phys.\  {\bf 42} (2001) 3127
  doi:10.1063/1.1372697
  [hep-th/0009181].

\bibitem{Malek:2013sp}
  E.~Malek,
  ``Timelike U-dualities in Generalised Geometry,''
  JHEP {\bf 1311} (2013) 185
  doi:10.1007/JHEP11(2013)185
  [arXiv:1301.0543 [hep-th]].


\bibitem{Blair:2013gqa}
  C.~D.~A.~Blair, E.~Malek and J.~H.~Park,
  ``M-theory and Type IIB from a Duality Manifest Action,''
  JHEP {\bf 1401} (2014) 172
  doi:10.1007/JHEP01(2014)172
  [arXiv:1311.5109 [hep-th]].





\bibitem{Blumenhagen:2011ph}
  R.~Blumenhagen, A.~Deser, D.~Lust, E.~Plauschinn and F.~Rennecke,
  ``Non-geometric Fluxes, Asymmetric Strings and Nonassociative Geometry,''
  J.\ Phys.\ A {\bf 44} (2011) 385401
  doi:10.1088/1751-8113/44/38/385401
  [arXiv:1106.0316 [hep-th]].



\bibitem{Blumenhagen:2012nt}
  R.~Blumenhagen, A.~Deser, E.~Plauschinn and F.~Rennecke,
  ``Non-geometric strings, symplectic gravity and differential geometry of Lie algebroids,''
  JHEP {\bf 1302} (2013) 122
  doi:10.1007/JHEP02(2013)122
  [arXiv:1211.0030 [hep-th]].



\bibitem{Dibitetto:2012rk}
  G.~Dibitetto, J.~J.~Fernandez-Melgarejo, D.~Marques and D.~Roest,
  ``Duality orbits of non-geometric fluxes,''
  Fortsch.\ Phys.\  {\bf 60} (2012) 1123
  doi:10.1002/prop.201200078
  [arXiv:1203.6562 [hep-th]].


\bibitem{Cederwall:2014opa}
  M.~Cederwall,
  ``T-duality and non-geometric solutions from double geometry,''
  Fortsch.\ Phys.\  {\bf 62} (2014) 942
  doi:10.1002/prop.201400069
  [arXiv:1409.4463 [hep-th]].


\bibitem{Berkeley:2014nza}
  J.~Berkeley, D.~S.~Berman and F.~J.~Rudolph,
  ``Strings and Branes are Waves,''
  JHEP {\bf 1406} (2014) 006
  doi:10.1007/JHEP06(2014)006
  [arXiv:1403.7198 [hep-th]].

\bibitem{Berman:2014jsa}
  D.~S.~Berman and F.~J.~Rudolph,
  ``Branes are Waves and Monopoles,''
  JHEP {\bf 1505} (2015) 015
  doi:10.1007/JHEP05(2015)015
  [arXiv:1409.6314 [hep-th]].
  
\bibitem{Lee:2016qwn}
  K.~Lee, S.~J.~Rey and Y.~Sakatani,
  ``Effective Action for Non-Geometric Fluxes from Duality Covariant Actions,''
  arXiv:1612.08738 [hep-th].



\bibitem{Choi:2015bga}
  K.~S.~Choi and J.~H.~Park,
  ``Standard Model as a Double Field Theory,''
  Phys.\ Rev.\ Lett.\  {\bf 115} (2015) no.17,  171603
  doi:10.1103/PhysRevLett.115.171603
  [arXiv:1506.05277 [hep-th]].



\bibitem{Bekaert:2016isw}
  X.~Bekaert and J.~H.~Park,
  ``Higher Spin Double Field Theory : A Proposal,''
  JHEP {\bf 1607} (2016) 062
  doi:10.1007/JHEP07(2016)062
  [arXiv:1605.00403 [hep-th]].



\bibitem{Hohm:2015ugy}
  O.~Hohm and D.~Marques,
  ``Perturbative Double Field Theory on General Backgrounds,''
  Phys.\ Rev.\ D {\bf 93} (2016) no.2,  025032
  doi:10.1103/PhysRevD.93.025032
  [arXiv:1512.02658 [hep-th]].



\bibitem{Park:2015bza}
  J.~H.~Park, S.~J.~Rey, W.~Rim and Y.~Sakatani,
  ``$\ODD$ covariant Noether currents and global charges in double field theory,''
  JHEP {\bf 1511} (2015) 131
  doi:10.1007/JHEP11(2015)131
  [arXiv:1507.07545 [hep-th]].
  
  
\bibitem{Blair:2015eba}
  C.~D.~A.~Blair,
  ``Conserved Currents of Double Field Theory,''
  JHEP {\bf 1604} (2016) 180
  doi:10.1007/JHEP04(2016)180
  [arXiv:1507.07541 [hep-th]].

\bibitem{Jeon:2012hp}
  I.~Jeon, K.~Lee, J.~H.~Park and Y.~Suh,
  ``Stringy Unification of Type IIA and IIB Supergravities under N=2 D=10 Supersymmetric Double Field Theory,''
  Phys.\ Lett.\ B {\bf 723} (2013) 245
  doi:10.1016/j.physletb.2013.05.016
  [arXiv:1210.5078 [hep-th]].



\bibitem{Grana:2005jc}
  M.~Grana,
  ``Flux compactifications in string theory: A Comprehensive review,''
  Phys.\ Rept.\  {\bf 423} (2006) 91
  doi:10.1016/j.physrep.2005.10.008
  [hep-th/0509003].



\bibitem{Son:2013rqa}
  D.~T.~Son,
  ``Newton-Cartan Geometry and the Quantum Hall Effect,''
  arXiv:1306.0638 [cond-mat.mes-hall].




 

    
 \bibitem{BergshoeffSimons}
 E.~Bergshoeff,  ``Applied Newton-Cartan Geometry," review talk at Simons Center, 
  \url{http://scgp.stonybrook.edu/video_portal/video.php?id=3051}.
 
 
\bibitem{Andringa:2012uz}
  R.~Andringa, E.~Bergshoeff, J.~Gomis and M.~de Roo,
  ``'Stringy' Newton-Cartan Gravity,''
  Class.\ Quant.\ Grav.\  {\bf 29} (2012) 235020
  doi:10.1088/0264-9381/29/23/235020
  [arXiv:1206.5176 [hep-th]].

\bibitem{Harmark:2017rpg}
  T.~Harmark, J.~Hartong and N.~A.~Obers,
  ``Nonrelativistic strings and limits of the AdS/CFT correspondence,''
  Phys.\ Rev.\ D {\bf 96} (2017) no.8,  086019
  doi:10.1103/PhysRevD.96.086019
  [arXiv:1705.03535 [hep-th]].

\bibitem{Ko:2016dxa}
  S.~M.~Ko, J.~H.~Park and M.~Suh,
  ``The rotation curve of a point particle in stringy gravity,''
  JCAP {\bf 1706} (2017) no.06,  002
  doi:10.1088/1475-7516/2017/06/002
  [arXiv:1606.09307 [hep-th]].



\bibitem{Bergshoeff:2017btm}
  E.~Bergshoeff, J.~Gomis, B.~Rollier, J.~Rosseel and T.~ter Veldhuis,
  ``Carroll versus Galilei Gravity,''
  JHEP {\bf 1703} (2017) 165
  doi:10.1007/JHEP03(2017)165
  [arXiv:1701.06156 [hep-th]].



\bibitem{Jeon:2010rw}
  I.~Jeon, K.~Lee and J.~H.~Park,
  ``Differential geometry with a projection: Application to double field theory,''
  JHEP {\bf 1104} (2011) 014
  doi:10.1007/JHEP04(2011)014
  [arXiv:1011.1324 [hep-th]].



\bibitem{Jeon:2011kp}
  I.~Jeon, K.~Lee and J.~H.~Park,
  ``Double field formulation of Yang-Mills theory,''
  Phys.\ Lett.\ B {\bf 701} (2011) 260
  doi:10.1016/j.physletb.2011.05.051
  [arXiv:1102.0419 [hep-th]].






\bibitem{Andriot:2013xca}
  D.~Andriot and A.~Betz,
  ``$\beta$-supergravity: a ten-dimensional theory with non-geometric fluxes, and its geometric framework,''
  JHEP {\bf 1312} (2013) 083
  doi:10.1007/JHEP12(2013)083
  [arXiv:1306.4381 [hep-th]].



\bibitem{Dabholkar:1990yf}
  A.~Dabholkar, G.~W.~Gibbons, J.~A.~Harvey and F.~Ruiz Ruiz,
  ``Superstrings and Solitons,''
  Nucl.\ Phys.\ B {\bf 340} (1990) 33.
  doi:10.1016/0550-3213(90)90157-9




    
 \bibitem{JHPBanff}
 J.~H.~Park,   ``Green-Schwarz superstring and Stringy Gravity in doubled-yet-gauged spacetime," talk at Banff International Research Station, 
  \url{http://www.birs.ca/events/2017/5-day-workshops/17w5018/videos/watch/201701231429-Park.html}.




\bibitem{Jeon:2011vx}
  I.~Jeon, K.~Lee and J.~H.~Park,
  ``Incorporation of fermions into double field theory,''
  JHEP {\bf 1111} (2011) 025
  doi:10.1007/JHEP11(2011)025
  [arXiv:1109.2035 [hep-th]].



\bibitem{Jeon:2012kd}
  I.~Jeon, K.~Lee and J.~H.~Park,
  ``Ramond-Ramond Cohomology and $\ODD$ T-duality,''
  JHEP {\bf 1209} (2012) 079
  doi:10.1007/JHEP09(2012)079
  [arXiv:1206.3478 [hep-th]].



\bibitem{Jeon:2011sq}
  I.~Jeon, K.~Lee and J.~H.~Park,
  ``Supersymmetric Double Field Theory: Stringy Reformulation of Supergravity,''
  Phys.\ Rev.\ D {\bf 85} (2012) 081501
   Erratum: [Phys.\ Rev.\ D {\bf 86} (2012) 089903]
  doi:10.1103/PhysRevD.86.089903, 10.1103/PhysRevD.85.081501, 10.1103/PhysRevD.85.089908
  [arXiv:1112.0069 [hep-th]].



\bibitem{Cho:2015lha}
  W.~Cho, J.~J.~Fernández-Melgarejo, I.~Jeon and J.~H.~Park,
  ``Supersymmetric gauged double field theory: systematic derivation by virtue of twist,''
  JHEP {\bf 1508} (2015) 084
  doi:10.1007/JHEP08(2015)084
  [arXiv:1505.01301 [hep-th]].



\bibitem{Duff:1986ne}
  M.~J.~Duff,
  ``Hidden String Symmetries?,''
  Phys.\ Lett.\ B {\bf 173} (1986) 289.
  doi:10.1016/0370-2693(86)90519-8



\bibitem{Hohm:2014sxa}
  O.~Hohm, A.~Sen and B.~Zwiebach,
  ``Heterotic Effective Action and Duality Symmetries Revisited,''
  JHEP {\bf 1502} (2015) 079
  doi:10.1007/JHEP02(2015)079
  [arXiv:1411.5696 [hep-th]].



\bibitem{Geissbuhler:2011mx}
  D.~Geissbuhler,
  ``Double Field Theory and N=4 Gauged Supergravity,''
  JHEP {\bf 1111} (2011) 116
  doi:10.1007/JHEP11(2011)116
  [arXiv:1109.4280 [hep-th]].



\bibitem{Aldazabal:2011nj}
  G.~Aldazabal, W.~Baron, D.~Marques and C.~Nunez,
  ``The effective action of Double Field Theory,''
  JHEP {\bf 1111} (2011) 052
   Erratum: [JHEP {\bf 1111} (2011) 109]
  doi:10.1007/JHEP11(2011)052, 10.1007/JHEP11(2011)109
  [arXiv:1109.0290 [hep-th]].



\bibitem{Grana:2012rr}
  M.~Grana and D.~Marques,
  ``Gauged Double Field Theory,''
  JHEP {\bf 1204} (2012) 020
  doi:10.1007/JHEP04(2012)020
  [arXiv:1201.2924 [hep-th]].



\bibitem{Geissbuhler:2013uka}
  D.~Geissbuhler, D.~Marques, C.~Nunez and V.~Penas,
  ``Exploring Double Field Theory,''
  JHEP {\bf 1306} (2013) 101
  doi:10.1007/JHEP06(2013)101
  [arXiv:1304.1472 [hep-th]].

\bibitem{Berman:2013cli}
  D.~S.~Berman and K.~Lee,
  ``Supersymmetry for Gauged Double Field Theory and Generalised Scherk-Schwarz Reductions,''
  Nucl.\ Phys.\ B {\bf 881} (2014) 369
  doi:10.1016/j.nuclphysb.2014.02.015
  [arXiv:1305.2747 [hep-th]].




\bibitem{Malek:2016vsh}
  E.~Malek,
  ``From Exceptional Field Theory to Heterotic Double Field Theory via K3,''
  JHEP {\bf 1703} (2017) 057
  doi:10.1007/JHEP03(2017)057
  [arXiv:1612.01990 [hep-th]];\\
  E.~Malek,
  ``7-dimensional ${\cal N}=2$ Consistent Truncations using $\mathrm{SL}(5)$ Exceptional Field Theory,''
  JHEP {\bf 1706} (2017) 026
  doi:10.1007/JHEP06(2017)026
  [arXiv:1612.01692 [hep-th]].




\bibitem{Malek:2017njj}
  E.~Malek,
  ``Half-maximal supersymmetry from exceptional field theory,''
  arXiv:1707.00714 [hep-th].



\bibitem{PREP}
J.~H.~Park and M.~Yata, \textit{in preparation.}


\bibitem{Hitchin:2004ut}
  N.~Hitchin,
  ``Generalized Calabi-Yau manifolds,''
  Quart.\ J.\ Math.\  {\bf 54} (2003) 281
  doi:10.1093/qjmath/54.3.281
  [math/0209099 [math-dg]].



\bibitem{Gualtieri:2003dx}
  M.~Gualtieri,
  ``Generalized complex geometry,''
  math/0401221 [math-dg].



\bibitem{Hitchin:2010qz}
  N.~Hitchin,
  ``Lectures on generalized geometry,''
  arXiv:1008.0973 [math.DG].



\bibitem{Coimbra:2011nw}
  A.~Coimbra, C.~Strickland-Constable and D.~Waldram,
  ``Supergravity as Generalised Geometry I: Type II Theories,''
  JHEP {\bf 1111} (2011) 091
  doi:10.1007/JHEP11(2011)091
  [arXiv:1107.1733 [hep-th]].



\bibitem{Coimbra:2012yy}
  A.~Coimbra, C.~Strickland-Constable and D.~Waldram,
  ``Generalised Geometry and type II Supergravity,''
  Fortsch.\ Phys.\  {\bf 60} (2012) 982
  doi:10.1002/prop.201100096
  [arXiv:1202.3170 [hep-th]].



\bibitem{Garcia-Fernandez:2013gja}
  M.~Garcia-Fernandez,
  ``Torsion-free generalized connections and Heterotic Supergravity,''
  Commun.\ Math.\ Phys.\  {\bf 332} (2014) no.1,  89
  doi:10.1007/s00220-014-2143-5
  [arXiv:1304.4294 [math.DG]].



\bibitem{Siegel:2015axg}
  W.~Siegel,
  ``Amplitudes for left-handed strings,''
  arXiv:1512.02569 [hep-th].



\bibitem{Hohm:2013jaa}
  O.~Hohm, W.~Siegel and B.~Zwiebach,
  ``Doubled $\alpha'$-geometry,''
  JHEP {\bf 1402} (2014) 065
  doi:10.1007/JHEP02(2014)065
  [arXiv:1306.2970 [hep-th]].



\bibitem{Huang:2016bdd}
  Y.~t.~Huang, W.~Siegel and E.~Y.~Yuan,
  ``Factorization of Chiral String Amplitudes,''
  JHEP {\bf 1609} (2016) 101
  doi:10.1007/JHEP09(2016)101
  [arXiv:1603.02588 [hep-th]].



\bibitem{Bergshoeff:2014jla}
  E.~Bergshoeff, J.~Gomis and G.~Longhi,
  ``Dynamics of Carroll Particles,''
  Class.\ Quant.\ Grav.\  {\bf 31} (2014) no.20,  205009
  doi:10.1088/0264-9381/31/20/205009
  [arXiv:1405.2264 [hep-th]].



\bibitem{Duval:2014uoa}
  C.~Duval, G.~W.~Gibbons, P.~A.~Horvathy and P.~M.~Zhang,
  ``Carroll versus Newton and Galilei: two dual non-Einsteinian concepts of time,''
  Class.\ Quant.\ Grav.\  {\bf 31} (2014) 085016
  doi:10.1088/0264-9381/31/8/085016
  [arXiv:1402.0657 [gr-qc]].



\bibitem{Bekaert:2015xua}
  X.~Bekaert and K.~Morand,
  ``Connections and dynamical trajectories in generalised Newton-Cartan gravity II. An ambient perspective,''
  arXiv:1505.03739 [hep-th].



\bibitem{Bernal:2002ph}
  A.~N.~Bernal and M.~Sanchez,
  ``Leibnizian, Galilean and Newtonian structures of space-time,''
  J.\ Math.\ Phys.\  {\bf 44} (2003) 1129
  doi:10.1063/1.1541120
  [gr-qc/0211030].



\bibitem{Bekaert:2014bwa}
  X.~Bekaert and K.~Morand,
  ``Connections and dynamical trajectories in generalised Newton-Cartan gravity I. An intrinsic view,''
  J.\ Math.\ Phys.\  {\bf 57} (2016) no.2,  022507
  doi:10.1063/1.4937445
  [arXiv:1412.8212 [hep-th]].



\bibitem{Duval:1993pe}
  C.~Duval,
  ``On Galilean isometries,''
  Class.\ Quant.\ Grav.\  {\bf 10} (1993) 2217
  doi:10.1088/0264-9381/10/11/006
  [arXiv:0903.1641 [math-ph]].



\bibitem{Kuenzle:1972zw}
  H.~P.~K\"unzle,
  ``Galilei and Lorentz structures on space-time - comparison of the corresponding geometry and physics,''
  Ann.\ Inst.\ H.\ Poincare Phys.\ Theor.\  {\bf 17} (1972) 337.



\bibitem{Duval:1984cj}
  C.~Duval, G.~Burdet, H.~P.~Kunzle and M.~Perrin,
  ``Bargmann Structures and Newton-Cartan Theory,''
  Phys.\ Rev.\ D {\bf 31} (1985) 1841.
  doi:10.1103/PhysRevD.31.1841



\bibitem{Duval:1990hj}
  C.~Duval, G.~W.~Gibbons and P.~Horvathy,
  ``Celestial mechanics, conformal structures and gravitational waves,''
  Phys.\ Rev.\ D {\bf 43} (1991) 3907
  doi:10.1103/PhysRevD.43.3907
  [hep-th/0512188].



\bibitem{Julia:1994bs}
  B.~Julia and H.~Nicolai,
  ``Null Killing vector dimensional reduction and Galilean geometrodynamics,''
  Nucl.\ Phys.\ B {\bf 439} (1995) 291
  doi:10.1016/0550-3213(94)00584-2
  [hep-th/9412002].



\bibitem{Bekaert:2013fta}
  X.~Bekaert and K.~Morand,
  ``Embedding nonrelativistic physics inside a gravitational wave,''
  Phys.\ Rev.\ D {\bf 88} (2013) no.6,  063008
  doi:10.1103/PhysRevD.88.063008
  [arXiv:1307.6263 [hep-th]].

\bibitem{Morand2017}
K.~Morand,
  ``{Connections and dynamical trajectories in generalised Newton-Cartan gravity III. A Platonic allegory},''
\textit{in preparation.}

\bibitem{Minguzzi:2006wz}
  E.~Minguzzi,
  ``Classical aspects of lightlike dimensional reduction,''
  Class.\ Quant.\ Grav.\  {\bf 23} (2006) 7085
  doi:10.1088/0264-9381/23/23/029
  [gr-qc/0610011].


\bibitem{BekaertSimons}
 X.~Bekaert,  ``Connections in Newton-Cartan Geometry: Intrinsic and Ambient Approaches," talk at Simons Center, 
\url{http://scgp.stonybrook.edu/video_portal/video.php?id=3058}.








  


\end{thebibliography}
\end{document}